\documentclass[superscriptaddress,nofootinbib,rmp]{revtex4}
\pdfoutput=1
\usepackage{aeguill}
\usepackage[pdftex]{graphicx}
\DeclareGraphicsExtensions{.pdf,.jpg,.png}
\usepackage[pdftex,colorlinks]{hyperref}

\newcommand{\Prob}{\ensuremath{\mathbb{P}}}
\newcommand{\Expect}[1]{\ensuremath{\mathbb{E}\left[ #1 \right]}}
\newcommand{\TimeAvg}[1]{\ensuremath{\left\langle #1 \right\rangle}}

\newcommand{\PastT}[1]{X^{#1}_{-\infty}}
\newcommand{\FutureT}[1]{X^{\infty}_{#1}}
\newcommand{\Lmax}{\Lambda}

\usepackage{amsfonts,amssymb,amsmath}
\usepackage{natbib}

\begin{document}
\title{The Computational Structure of Spike Trains}

\author{Robert Haslinger}
\affiliation{Martinos Center for Biomedical Imaging,Massachusetts General
  Hospital, Charlestown MA}
\affiliation{Department of Brain and Cognitive Sciences, Massachusetts 
Institute of Technology, Cambridge MA}
\thanks{}

\author{Kristina Lisa Klinkner} \affiliation{Department of Statistics, Carnegie Mellon
University, Pittsburgh PA}
\thanks{}

\author{Cosma Rohilla Shalizi} \affiliation{Department of Statistics, Carnegie
  Mellon University, Pittsburgh PA}
\affiliation{Santa Fe Institute, Santa Fe NM}
 \thanks{}

\date{September 2008; January 2009}

\begin{abstract}
\normalsize
  Neurons perform computations, and convey the results of those computations
  through the statistical structure of their output spike trains.  Here we
  present a practical method, grounded in the information-theoretic analysis of
  prediction, for inferring a minimal representation of that structure and for
  characterizing its complexity.  Starting from spike trains, our approach
  finds their {\em causal state models} (CSMs), the minimal hidden Markov
  models or stochastic automata capable of generating statistically-identical
  time series.  We then use these CSMs to objectively quantify both the
  generalizable structure and the idiosyncratic randomness of the spike train.
  Specifically, we show that the expected algorithmic information content (the
  information needed to describe the spike train exactly) can be split into
  three parts describing (1) the time-invariant structure (complexity) of the
  minimal spike-generating process, which describes the spike train {\em
    statistically}, (2) the randomness (internal entropy rate) of the minimal
  spike-generating process, and (3) a residual pure noise term not described by
  the minimal spike generating process.  We use CSMs to approximate each of
  these quantities.  The CSMs are inferred non-parametrically from the data,
  making only mild regularity assumptions, via the Causal State Splitting
  Reconstruction (CSSR) algorithm.  The methods presented here complement more
  traditional spike train analyses by describing not only spiking probability,
  and spike train entropy, but also the complexity of a spike train's
  structure.  We demonstrate our approach using both simulated spike trains and
  experimental data recorded in rat barrel cortex during vibrissa stimulation.
\end{abstract}

\maketitle

\section{Introduction}

The recognition that neurons are computational devices is one of the
foundations of modern neuroscience \citep{McCulloch-Pitts-immanent-calculus}.
However, determining the functional form of such computation is extremely
difficult, if only because while one often knows the output (the spikes) the
input (synaptic activity) is almost always unknown.  Often, therefore,
scientists must draw inferences about the computation from its results, namely
the output spike trains and their statistics.  In this vein, many researchers
have used information theory to determine, via calculation of the entropy rate,
a neuron's channel capacity, i.e., how much information the neuron could
conceivably transmit, given the distribution of observed spikes
\citep{Spikes-book}.  However, entropy quantifies randomness, and says little
about how much {\em structure} a spike train has, or the amount and type of
computation which must have, at a minimum, taken place to produce this
structure.  Here, and throughout this paper, we mean ``computational
structure'' information-theoretically, i.e., the most compact effective
description of a process capable of {\em statistically} reproducing the
observed spike trains.  The complexity of this structure is the number of bits
needed to describe it.  This is different from the algorithmic information
content of a spike train, which is the number of bits needed to reproduce the
latter {\em exactly}, describing not only its regularities, but also its
accidental, noisy details.

Our goal is to develop rigorous yet practical methods for determining the
minimal computational structure necessary and sufficient to generate neural
spike trains.  We are able to do this through non-parametric analysis of the
directly-observable spike trains, without resorting to {\em a priori}
assumptions about what kind of structure they have.  We do this by identifying
the minimal hidden Markov model (HMM) which can statistically predict the
future of the spike train without loss of information.  This HMM also generates
spike trains with the same statistics as the observed train.  It thus defines a
program which describes the spike train's computational structure, letting us
quantify, in bits, the structure's complexity.

From multiple directions, several groups, including our own, have shown that
minimal generative models of time series can be discovered by clustering
histories into ``states'', based on their conditional distributions over future
events
\citep{Knight-predictive-view,Grassberger-1986,Inferring-stat-compl,CMPPSS,
  Jaeger-operator-models,predictive-representations-of-state}.  The observed
time series need {\em not} be Markovian (few spike trains are), but the
construction always yields the minimal HMM capable of generating and predicting
the original process.  Following \textcite{CRS-thesis,CMPPSS}, we will call
such a HMM a ``Causal State Model'' (CSM).  Within this framework, the model
discovery algorithm called {\em Causal State Splitting Reconstruction}, or CSSR
\citep{CSSR-for-UAI} is an adaptive non-parametric method which consistently
estimates a system's CSM from time-series data. In this paper we adapt CSSR for
use in spike train analysis.

CSSR provides us with non-parametric estimates of the time- and history-
dependent spiking probabilities found by more familiar parametric analyses.
Unlike those analyses, it is also capable, in the limit of infinite data, of
capturing {\em all} the information about the computational structure of the
spike-generating process contained in the spikes themselves.  In particular,
the CSM quantifies the {\em complexity} of the spike-generating process by
showing how much information about the history of the spikes is relevant to
their future, i.e., how much information is needed to reproduce the spike train
statistically.  This is equivalent to the log of the effective number of
statistically-distinct states of the process
\citep{Grassberger-1986,Inferring-stat-compl,CMPPSS}.  While this is not the
same as the algorithmic information content, we show that CSMs can also
approximate the average algorithmic information content, splitting it into
three parts: (1) The generative process's complexity in our sense; (2) the {\em
  internal entropy rate} of the generative process, the extra information
needed to describe the exact state transitions the undergone while generating
the spike train; and (3) the {\em residual randomness} in the spikes,
unconstrained by the generative process.  The first of these quantifies the
spike train's structure, the last two its randomness.

Below, we give precise definitions of these quantities, both their ensemble
averages (\S \ref{subsect:alginfocont}) and their functional dependence on time
(\S \ref{sec:time-varying-compl}).  The time-dependent versions allow us to
determine when the neuron is traversing states requiring complex
descriptions. Our methods put hard numerical lower bounds on the amount of
computational structure which must be present to generate the observed spikes.
They also quantify, in bits, the extent to which the neuron is driven by
external forces.  We demonstrate our approach using both simulated and
experimentally recorded single-neuron spike trains. We discuss the
interpretation of our measures, and how they add to our understanding of
neuronal computation.

\section{Theory and Methods}

Throughout this paper we treat spike trains as stochastic binary time series,
with time divided into discrete, equal-duration bins steps (typically at one
millisecond resolution); ``1'' corresponds to a spike and ``0'' to no spike.
Our aim is to find a minimal description of the computational structure present
in such a time series.  Heuristically, the structure present in a spike train
can be described by a ``program'' which can reproduce the spikes
statistically. The information needed to describe this program (loosely
speaking the program length) quantifies the structure's complexity.  Our
approach uses minimal, optimally predictive HMMs, or {\em Causal State Models}
(CSMs), reconstructed from the data, to describe the program.  (We clarify our
use of ``minimal'' below.)  The CSMs are then used to calculate various
measures of the computational structure, such as its complexity.

The states are chosen so that they are optimal predictors of the spike train's
future, using only the information available from the train's history.  (We
discuss the limitations of this below.)  Specifically the states $S_t$ are
defined by grouping the histories of past spiking activity $X_{-\infty}^t$
which occur in the spike train into equivalence classes, where all members of a
given equivalence class are statistically equivalent in terms of predicting the
future spiking $X_{t+1}^{\infty}$.  ($X_{t'}^t$ denotes the sequence of random
observables, i.e., spikes or their absence, between $t'$ and $t>t'$ while $X_t$
denotes the random observable at time $t$. The notation is similar for the
states.)  This construction ensures that the causal states are Markovian, even
if the spike train is not \cite[Lemma 6, p.\ 839]{CMPPSS}.  Therefore, at all
times $t$ the system and its possible future evolution(s) can be specified by
the state $S_t$.  Like all HMMs, a CSM can be represented pictorially by a
directed graph, with nodes standing for the process's hidden states and
directed edges the possible transitions between these states.  Each edge is
labeled with the observable/symbol emitted during the corresponding transition
(``1'' for a spike and ``0'' for no spike), and the probability of traversing
that edge given that the system started in that state.  The CSM also specifies
the time-averaged probability of occupying any state (via the ergodic theorem
for Markov chains).

\begin{figure}[p]
\includegraphics{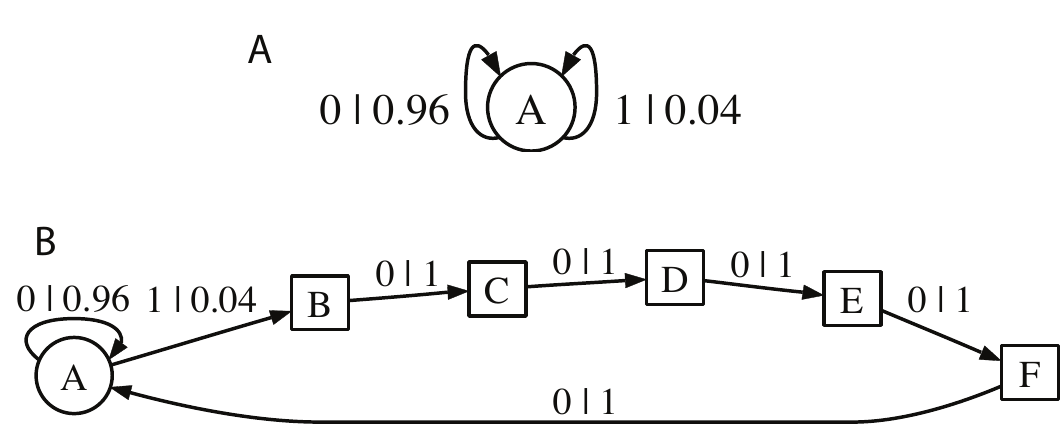}
\caption{Two simple CSMs reconstructed from 200 sec of simulated spikes
  using CSSR.  States are represented as the nodes of a directed graph.  The
  transitions between states are labeled with the symbol emitted during the
  transition (1 = spike, 0 = no spike) and the probability of the transition
  given the origin state.  (A) The CSM for a 40 Hz Bernoulli spiking process
  consists of a single state $A$ which always transitions back to itself,
  emitting a spike with probability $p=0.04$ per msec.  (B) CSM for 40 Hz
  Bernoulli spiking process with a 5 msec refractory period imposed after
  each spike.  State $A$ again spikes with probability $p=0.04$.  Upon
  spiking the CSM transitions through a deterministic chain of states
  $B$--$F$ (squares) which represent the refractory period.  The increased
  structure of the refractory period requires a more complex representation.}
\label{fig:1}
\end{figure}

The theory is described in more detail below, but at this point examples may
clarify the ideas.  Figures \ref{fig:1} A and B show two simple CSMs.  Both are
built from simulated $\approx 40$ Hz spike trains 200 seconds in length (1 msec
time bins, $p=0.04$ IID at each time when spiking is possible).  However, spike
trains generated from the CSM in Figure \ref{fig:1} B have a 5 msec refractory
period after each spike (when $p=0$), while the spiking rate in non-refractory
periods is still 40 Hz ($p=0.04$).  The refractory period is additional
structure, represented by the extra states.  State $A$ represents the status of
the neuron during 40 Hz spiking, outside of the refractory periods.  While in
this state, the neuron either emits no spike ($X_{t+1}=0$), staying in state
$A$, or emits a spike ($X_{t+1}=1$) with probability $p=0.04$ and moves to
state $B$.  The equivalence class of past spiking histories defining state $A$
therefore includes all past spiking histories for which the most recent five
symbols are 0, symbolically $\{*00000\}$.  State $B$ is the neuron's state
during the first msec of the refractory period.  It is defined by the set of
spiking histories $\{*1\}$.  No spike can be emitted during a refractory period
so the transition to state $C$ is certain and the symbol emitted is always '0'.
In this manner the neuron proceeds through states $C$ to $F$ and back to state
$A$ whereupon it is possible to spike again.

The rest of this section is divided into four subsections.  First, we briefly
review the formal theory behind CSMs (for details, see
\textcite{CRS-thesis,CMPPSS}) and discuss why they can be considered a good
choice for understanding the structural content of spike trains.  Second, we
describe the {\em Causal State Splitting Reconstruction (CSSR)} algorithm used
to reconstruct CSMs from observed spike trains \citep{CSSR-for-UAI}.  We
emphasize that CSSR requires no {\em a priori} knowledge of the structure of
the CSM which is discovered from the spike train.  Third, we discuss two
different notions of spike train structure, namely statistical complexity and
algorithmic information content.  These two measures can be interpreted as
different aspects of a spike train's computational structure, and each can be
related to the reconstructed CSM.  Fourth and finally, we show how the
reconstructed CSM can be used to predict spiking, measure the neural response
and detect the influence of external stimuli.

\subsection{Causal State Models}
\label{subsect:CSMtheory}

The foundation of the theory of causal states is the concept of a {\em
  predictively sufficient statistics}.  A statistic, $\eta$, on one random
variable, $X$, is sufficient for predicting another random variable, $Y$, when
$\eta(X)$ and $X$ have the same information\footnote{\normalsize See
  \textcite{Cover-and-Thomas} for information-theoretic definitions and
  notation.} about $Y$, $I[X;Y] = I[\eta(X);Y]$.  This holds if and only if $X$
and $Y$ are conditionally independent given $\eta(X)$: $\Prob(Y|X,\eta(X)) =
\Prob(Y|\eta(X))$.  This is a close relative of the familiar idea of {\em
  parametric} sufficiency; in Bayesian statistics, where parameters are random
variables, parametric sufficiency is a special case of predictive sufficiency
\citep{Bernardo-et-al-bayesian-theory}.  Predictive sufficiency shares all of
parametric sufficiency's optimality properties
\citep{Bernardo-et-al-bayesian-theory}.  However, a statistic's predictive
sufficiency depends only on the actual joint distribution of $X$ and $Y$, not
on any parametric model of that distribution.  Again as in the parametric case,
a {\em minimal} predictively sufficient statistic $\epsilon$ is one which is a
function of every other sufficient statistic $\eta$, i.e., $\epsilon(X) =
h(\eta(X))$ for some $h$.  Minimal sufficient statistics are the most compact
summaries of the data which retain all the predictively-relevant information.
A basic result is that a minimal sufficient statistic always exists and is
(essentially) unique, up to isomorphism
\citep{Bernardo-et-al-bayesian-theory,CMPPSS}.

In the context of stochastic processes, such as spike trains, $\epsilon$ is the
minimal sufficient statistic of the history $X^t_{-\infty}$ for predicting
future of the process, $X^{\infty}_{t+1}$.  This statistic is the optimal
predictor of the observations.  The sequence of values of the minimal
sufficient statistic, $S_t = \epsilon(X^t_{-\infty})$, is another stochastic
process.  This process is always a homogeneous Markov chain, whether or not the
$X_t$ process is \citep{Knight-predictive-view,CMPPSS}.  Turned around, this
means that the original $X_t$ process is always a random function of a
homogeneous Markov chain, whose latent states, named the {\em causal states} by
\textcite{Inferring-stat-compl}, are optimal, minimal predictors of the future
of the time series.

A {\em causal state model} or {\em causal state machine} is a stochastic
automaton or HMM constructed so that its Markov states are minimal sufficient
statistics for predicting the future of the spike train, and consequently can
generate spike trains statistically identical to those observed.
\footnote{\normalsize Some
  authors use ``hidden Markov Model'' only for models where the current
  observation is independent of all other variables given the current state,
  and call the broader class which includes CSMs ``partially observable Markov
  model''.}  Causal state {\em reconstruction} means inferring the causal
states from the observed spike train.  Following
\textcite{Inferring-stat-compl,CMPPSS}, the causal states can be seen as
equivalence classes of spike-train histories $X_{-\infty}^{t}$ which maximize
the mutual information between the state(s) and the future of the spike train
$X_{t+1}^{\infty}$.  Because they are sufficient, they predict the future of
the spike train as well as it can be predicted from its history alone.  Because
they are minimal, the number of states or equivalence classes is as small as it
can be without discarding predictive power.\footnote{\normalsize 
There may exist more
  compact representations, but then the states, or their equivalents, can {\em
    never} be empirically identified --- see \textcite[thm.\ 3, p.\
  846]{CMPPSS}, or \textcite{Lohr-Ay-generative-prediction}.}

Formally, two histories, $x^{-}$ and $y^{-}$, are equivalent when
$\Prob(\FutureT{t+1}|\PastT{t} = x^{-}) = \Prob(\FutureT{t+1}|\PastT{t} =
y^{-})$.  The equivalence class of $x^{-}$ is $[x^{-}]$.  Define the function
which maps histories to their equivalence classes:
\[
\lefteqn{\epsilon(x^{-}) \equiv [x^{-}]}
\]
\[
= \left\{y^{-} : \Prob(\FutureT{t+1}|\PastT{t} = y^{-}) = \Prob(\FutureT{t+1}|\PastT{t} = x^{-})\right\}
\]
The causal states are the possible values of $\epsilon$, i.e., the equivalence
classes; each corresponds to a distinct distribution for the future.  The state
at time $t$ is $S_t = \epsilon(\PastT{t})$.  Clearly, $\epsilon(x^{-})$ is a
sufficient statistic.  It is also minimal, since if $\eta$ is sufficient, then
$\eta(x^{-}) = \eta(y^{-})$ implies $\epsilon(x^{-}) = \epsilon(y^{-})$.  One
can further show \cite[Theorem 3]{CMPPSS} that $\epsilon$ is the {\em unique}
minimal sufficient statistic, meaning that any other must be isomorphic to it.

In addition to being minimal sufficient statistics, the causal states have some
other important properties which make them ideal for quantifying structure
\citep{CMPPSS}.  (1) As mentioned, $\left\{S_t\right\}$ is a Markov process,
and one can write the observed process $X$ as a random function of the causal
state process, i.e., $X$ has a natural hidden-Markov-model representation.  (2)
The causal states are recursively calculable; there is a function $T$ such that
$S_{t+1} = T(S_t,X_{t+1})$ --- see Appendix \ref{app:filtering}.
(3)  CSMs are closely related to the ``predictive state representations'' of
controlled dynamical systems \citep{predictive-representations-of-state}; see
Appendix \ref{sec:CSTs-and-PSRs}.

\subsection{Causal State Splitting Reconstruction}
\label{subsect:CSSR}

Our goal is to find a minimal sufficient statistic for the spike train, which
will form a hidden Markov model.  As stated previously, the states of this
model are equivalence classes of spiking histories $X_{{-\infty}}^t$.  In
practice, we need an algorithm which can both cluster histories into groups
which preserve their conditional distribution of futures, and find the history
length $\Lmax$ at which the past may be truncated while preserving the
computational structure of the spike train.  The former is accomplished by the
CSSR algorithm \citep{CSSR-for-UAI} for inferring causal states from data by
building a recursive next-step-sufficient statistic.\footnote{\normalsize A
  next-step-sufficient statistic contains all the information needed for
  optimal one-step-ahead prediction, $I[X_{t+1};\eta(X_{-\infty}^t)] =
  I[X_{t+1};X_{-\infty}^t]$, but not necessarily for longer predictions.  CSSR
  relies on the theorem that if $\eta$ is next-step sufficient, and it is
  recursively calculable, then $\eta$ is sufficient for the whole of the future
  \cite[pp.\ 842--843]{CMPPSS}.  CSSR first finds a next-step sufficient
  statistic, and then refines it to be recursive.}  We do the latter by
minimizing Schwartz's Bayesian Information Criterion (BIC) over $\Lmax$.

To save space, we just sketch the CSSR algorithm here.\footnote{\normalsize
In addition to
  \textcite{CSSR-for-UAI}, which gives pseudocode, some details of convergence,
  and applications to process classification, are treated in
  \textcite{CSSR-for-classification,CSSR-for-Ann-Stat}.  An open-source C++
  implementation is available at \url{http://bactra.org/CSSR/}.  The CSMs
  generated by CSSR can be displayed graphically, as we do in this paper, with
  the open-source program \texttt{dot} (\url{http://www.graphviz.org/}} CSSR
starts by treating the process as an independent, identically-distributed
sequence, with one causal state.  It adds states when statistical tests show
that the current set of states is not sufficient.  Suppose we have a sequence
$x_1^N = x_1, x_2, \ldots x_N$ of length $N$ from a finite alphabet ${\mathcal
  A}$ of size $k$.  We wish to derive from this an estimate $\hat{\epsilon}$ of
the minimal sufficient statistic $\epsilon$.  We do this by finding a set
$\Sigma$ of states, each of which will be a set of strings, or finite-length
histories.  The function $\hat{\epsilon}$ will then map a history $x^{-}$ to
whichever state contains a suffix of $x^{-}$ (taking ``suffix'' in the usual
string-manipulation sense).  Although each state can contain multiple suffixes,
one can check \citep{CSSR-for-UAI} that the mapping $\hat{\epsilon}$ will never
be ambiguous.

The \emph{null hypothesis} is that the process is Markovian on the basis of the
states in $\Sigma$,
\begin{equation}
\label{NullHyp}
\Prob(X_{t}|X^{t-1}_{t-L} = a {x}^{t-1}_{t-L+1})  = 
\Prob(X_{t}|\hat{S} = \hat{\epsilon}(x^{t-1}_{t-L+1})) 
\end{equation}
for all $a \in {\mathcal A}$.  In words, adding an extra piece of history does
not change the conditional distribution for the next observation.  We can check
this with standard statistical tests, such as $\chi^2$ or Kolmogorov-Smirnov.
In this paper, we used a KS test of size $\alpha=0.01$.\footnote{\normalsize 
For finite
  $N$, decreasing $\alpha$ tends to yield simpler CSMs with fewer states.  In a
  sense, it is a sort of regularization coefficient.  The influence of this
  regularization diminishes as $N$ increases.  For the data used in the Results
  section of this paper, varying $\alpha$ in the range $0.001<\alpha<0.1$ made
  little difference.}  If we reject this hypothesis, we fall back on a {\em
  restricted alternative hypothesis}, that we have the right set of conditional
distributions, but have matched them with the wrong histories.  That is,
\begin{equation}
\label{RestAltHyp}
\Prob(X_{t}|X^{t-1}_{t-L} = a{x}^{t-1}_{t-L+1}) = \Prob(X_{t}|\hat{S} = s^{*})
\end{equation}
for some $s^{*} \in \Sigma$, but $s^{*} \neq \hat{\epsilon}(x^{t-1}_{t-L+1})$.
If this hypothesis passes a test of size $\alpha$, then $s^{*}$ is the state to
which we assign the history\footnote{\normalsize If more than one 
such state $s^{*}$
  exists, we chose the one for which ${\widehat{\Prob}}(X_{t}|\hat{S} = s^{*})$
  differs least, in total variation distance, from
  ${\widehat{\Prob}}(X_{t}|^{t-1}_{t-L} = a {x}^{t-1}_{t-L+1})$, which is
  plausible and convenient.  However, which state we chose is irrelevant in the
  limit $N\rightarrow\infty$, so long as the difference between the
  distributions is not statistically significant.}.  Only if the
(\ref{RestAltHyp}) is itself rejected do we create a new state, with the suffix
$a{x}^{t-1}_{t-L+1}$.  \footnote{\normalsize 
The conceptually similar algorithm of
  \textcite{Kennel-Mees-context-tree-modeling} in effect always creates a new
  state, which leads to more complex models, sometimes infinitely more complex
  ones; see \textcite{CSSR-for-UAI}.}

The algorithm itself has three phases.  Phase I initializes $\Sigma$ to a
single state, which contains only the null suffix $\emptyset$.  (That is,
$\emptyset$ is a suffix of any string.)  The length of the longest suffix in
$\Sigma$ is $L$; this starts at 0.  Phase II iteratively tests the successive
versions of the null hypothesis, Eq.\ \ref{NullHyp}, and $L$ increases by one
each iteration, until we reach some maximum length $\Lmax$.  At the end of II,
$\hat{\epsilon}$ is (approximately) next-step sufficient.  Phase III makes
$\hat{\epsilon}$ recursively calculable, by splitting the states until they
have deterministic transitions.  Under mild technical conditions (a finite true
number of states, etc.), CSSR converges in probability on the correct CSM as
$N\rightarrow \infty$, provided only that $\Lmax$ is long enough to
discriminate all of the states.  The error of the predicted distributions of
futures $\Prob(X_{t+1}^{\infty}|X_{-\infty}^t)$, measured by total variation
distance, decays as $N^{-1/2}$.  Section 4 of \textcite{CSSR-for-UAI} details
CSSR's convergence properties. Comparisons of CSSR's performance
with that of more traditional 
expectation maximization based approaches can also be found in
\textcite{CSSR-for-UAI} as can time complexity bounds for the algorithm.
Depending upon the machine used, CSSR can process an $N=10^6$ time series
in under a minute.

\subsubsection{Choosing $\Lmax$}

CSSR requires no {\em a priori} knowledge of the CSM's structure, but does need
a choice of of $\Lmax$; here pick it by minimizing the BIC of the reconstructed
models over $\Lmax$, i.e.,
\begin{eqnarray}
BIC \equiv -2\log \mathcal{L}  + d\log{N}
\label{eqn:BIC-defined}
\end{eqnarray}
where $\mathcal{L}$ is the likelihood, $N$ is the data length and $d$ is the
number of model parameters, in our case the number of predictive states
\footnote{\normalsize 
The number of independent parameters $d$ involved in describing the
  CSM will be (number of states)*(number of symbols - 1) since the sum of the
  outgoing probabilities for each state is constrained to be 1.  Thus, for a
  binary alphabet, $d =$ number of states.}  BIC's logarithmic-with-$N$ penalty
term helps keep the number of causal states from growing too quickly with
increased data size, which is why we use it instead of the Akaike Information
Criterion (AIC).  Also, BIC is known to be consistent for selecting the order
of Markov chains and variable-length Markov models
\citep{Csiszar-Talata-context-trees}, both of which are sub-classes of CSMs.

Writing the observed spike train as $x_1^N$, and the state sequence as $s_0^N$,
the total likelihood of the spike train is
\begin{eqnarray}
\mathcal{L}= \sum_{s_0^N \in \Sigma^{N+1}}{\Prob(X_1^N=x_1^N|S_0^N = s_0^N) \Prob(S_0^N = s_0^N)}~,
\end{eqnarray}
the sum over all possible causal state sequences of the joint probability of
the spike train and the state sequence.  Since the states update recursively,
$s_{t+1} = T(s_t,x_{t+1})$, the starting state $s_0$ and the spike train
$x_1^N$ fix the entire state sequence $s_0^N$.  Thus the sum over state
sequences can be replaced by a sum over initial states
\begin{eqnarray}
\mathcal{L}= \sum_{s_i \in \Sigma}{\Prob(X_1^N=x_1^N|S_0 = s_i) \Prob(S_0 = s_i)}
\label{eqn:likelihood-as-sum-over-initial-states}
\end{eqnarray}
with the state probabilities $\Prob(S_0 = s_i)$ coming from the CSM.
By the Markov property,
\begin{eqnarray}
\Prob(X_1^N=x_1^N|S_0 = s_i) = \prod_{j=1}^{N}{\Prob(X_j =
x_j| S_{j-1}=s_{j-1})}
\label{eqn:likelihood-as-factor-over-sequence}
\end{eqnarray}

Selecting $\Lmax$ is now straightfoward: for each value of $\Lmax$, we build
the CSM from the spike train, calculate the likelihood using Eq.\
\ref{eqn:likelihood-as-sum-over-initial-states} and
\ref{eqn:likelihood-as-factor-over-sequence}, and pick the value, and CSM,
minimizing Eq.\ \ref{eqn:BIC-defined}.  We try all values of $\Lmax$ up to a
model-independent upper bound.  For a wide range of stochastic processes,
\textcite{Marton-Shields} showed that the length $m$ of subsequences for which
probabilities can be consistently and non-parametrically estimated can grow as
fast as $\log{N}/h$, where $h$ is the entropy rate, but no faster.  CSSR
estimates the distribution of the next symbol given the previous $\Lmax$
symbols, which is equivalent to estimating joint probabilities of blocks of
length $m=\Lmax+1$.  Thus Marton and Shield's result limits the 
usable values of $\Lmax$:
\begin{equation}
\label{eqn:marton-shields} \Lmax \leq \frac{\log{N}}{h} -1
\end{equation}
Using Eq.\ \ref{eqn:marton-shields} requires the entropy rate $h$.  The latter
can either be upper bounded as the log of the alphabet size (here, $\log{2} =
1$), or by some other, less pessimistic, estimator of the entropy rate (such as
the output of CSSR with $\Lmax = 1$).  Use of an upper bound on $h$ results in
a conservative maximum value for $\Lmax$.  For example, a 30 minute experiment
with 1 msec time bins lets us use {\em at least} $\Lmax\approx 20$ by the most
pessimistic estimate of $h=1$; the actual maximum value of $\Lmax$ may be much
larger.  We use $\Lmax \le 25$ in this paper but see no indication that this
can't be extended further, if need be.

\subsubsection{Condensing the CSM}
\label{section:condense}

For real neural data, the number of causal states can be very large ---
hundreds or more.  This creates an interpretation problem, if only because it
is hard to fit such a CSM on a single page for inspection.  We thus developed a
way to reduce the full CSM while still accounting for most of the spike train's
structure.  Our ``state culling'' technique found the least-probable states and
selectively removed them, appropriately redirecting state transitions and
reassigning state occupation probabilities.  By keeping the most probable
states, we focus on the ones which contribute the most to the spike train's
structure and complexity.  Again, we used BIC as our model selection criterion.

First, we sorted the states by probability, finding the least probable state
(``remove'' state) with a single incoming edge from a state (its ``ancestor'')
with outgoing transitions to two different states, the remove state and a
second, ``keep'' state.  We redirected both of the ancestor's outgoing edges to
the keep state.  Second, we reassigned the remove state's outgoing transitions
to the keep state.  If the outgoing transitions from the keep state were still
deterministic (at most a single 0 emitting edge and a single 1 emitting edge),
we stopped.  If the transitions were non-deterministic, we merged states
reached by emitting 0s with each other (likewise those reached by 1s),
repeating this until termination.  Third, we checked that there existed a state
sequence of the new model which could generate the observed spikes.  If there
was, we accepted the new CSM.  If not, we rejected the new CSM and chose the
next lowest probability state from the original CSM to remove.

This culling was iterated until removing any state made it impossible for the
CSM to generate the spike train.  At each iteration, we calculated BIC (as
described in the previous section), and ultimate chose the culled CSM with the
minimum BIC.  This gave a culled CSM for each value of $\Lmax$; the final one
we used was chosen after also minimizing BIC over $\Lmax$.  The CSMs shown
below in the results section paper result from this minimizing of BIC over
$\Lmax$ and state culling.

\subsubsection{ISI Bootstrapping}
\label{subsect:ISIbootstrap}

While we do model selection with BIC, we also want to do model checking or
adequacy-testing.  For the most part, we do this by using the CSM to bootstrap
point-wise confidence bounds on the interspike interval (ISI) distribution, and
checking their coverage of the empirical ISI distribution.  Because this
distribution is not used by CSSR in reconstructing the CSM, it provides a check
on the latter's ability to accurately describe the spike-train's statistics.

Specifically, we generated confidence bounds as follows.  To simulate one spike
train, we picked a random starting state according to the CSM's inferred
state-occupation probabilities, and then ran the CSM forward for $N$
time-steps, $N$ being the length of the original spike train.  This gives a
binary time-series, where a ``1'' stands for a spike and a ``0'' for no-spike,
and gave us a sample of inter-spike intervals from the CSM.  This in turn gave
an ``empirical'' ISI distribution.  Repeated over ${10}^4$ independent runs of
the CSM, and taking the $0.005$ and $0.995$ quantiles of the distributions at
each ISI length, gives 99\% pointwise confidence bounds.  (Pointwise bounds are
necessary because of the ISI distribution often modulates rapidly with ISI
length.)  If the CSM is correct, the empirical ISI will, by chance, lie outside
the bounds at $\approx1\%$ of the ISI lengths.

If we split the data into training and validation sets, a CSM reconstructed
from the training set can be used to bootstrap ISI confidence bounds, which can
be compared to the ISI distribution of the test set.  We discuss this sort of
of cross validation, as well as an additional test based on the time rescaling
theorem, in Appendix \ref{Appendix:Cross-Validate}.

\subsection{Complexity and Algorithmic Information Content}
\label{subsect:alginfocont}

The {\em algorithmic information content} $K(x_1^n)$ of a sequence $x_1^n$ is
the length of the shortest complete (input-free) computer program which will
output $x_1^n$ exactly and then halt \citep{Cover-and-Thomas}
\footnote{\normalsize The
  algorithmic information content is also called the Kolmogorov complexity.  We
  do not use this term, to avoid confusion with our ``complexity'' $C$the
  information needed to reproduce the spike train {\em statistically} rather
  than {\em exactly} (Eq.\ \ref{eqn:C-defined}).  See \textcite{Badii-Politi}
  for a detailed comparison of complexity measures.}.  In general, $K(x_1^n)$
is strictly uncomputable, but when $x_1^n$ is the realization of a stochastic
process $X_1^n$, the ensemble-averaged algorithmic information essentially
coincides with the Shannon entropy (``Brudno's theorem''; see
\textcite{Badii-Politi}), reflecting the fact that both are maximized for
completely random sequences \citep{Cover-and-Thomas}.  Both the algorithmic
information and the Shannon entropy can be conveniently written in terms of a
minimal sufficient statistic $Q$:
\begin{eqnarray}
\Expect{K(X_1^n)} &=& H[X_1^n] + o(n) \nonumber\\
&=& H[Q] + H[X_1^n|Q] + o(n)
\label{eq:algorithmicinfo}
\end{eqnarray}
The equality $H[X_1^n] = H[Q] + H[X_1^n|Q]$ holds because $Q$ is a function
of $X_1^n$, so $H[Q|X_1^n] = 0$.

The key to determining a spike train's expected algorithmic information is thus
to find a minimal sufficient statistic.  By construction, causal state models
provide exactly this; a minimal sufficient statistic for $x_1^n$ is the state
sequence $s_0^n = s_0, s_1, \ldots s_n$ \citep{CMPPSS}.  Thus the
ensemble-averaged algorithmic information content, dropping terms $o(n)$ and
smaller, is
\begin{eqnarray}
\Expect{K(X_1^n)} &=& 
H[S_0^n] + H[X_1^n|S_0^n] \nonumber\\
&=& H[S_0] + \sum_{i=1}^n{H[S_i|S_{i-1}] + \sum_{i=1}^n H[X_i|S_i,S_{i-1}]}
\label{jointHsplit}
\end{eqnarray}
Going from the first to the second line uses the causal states' Markov
property.  Assuming stationarity, Eq.\ \ref{jointHsplit} becomes
\begin{eqnarray}
  \Expect{K(X_1^n)} &=& H[S_t] + n \left( H[S_{t}|S_{t-1}] + H[X_t | S_t, S_{t-1}] \right)
  \nonumber\\
  &=& C + n\left( J + R \right)
\label{eqn:final-decomp-of-average-complexity}
\end{eqnarray}
This separates terms representing structure from those representing randomness.

The first term in Eq.\ \ref{eqn:final-decomp-of-average-complexity} is the {\em
  complexity}, $C$, of the spike-generating process
\citep{Grassberger-1986,Inferring-stat-compl,QSO-in-PRL}.
\begin{equation}
C = H[S_t] = -\Expect{\log{\Prob(S_t)}}
\label{eqn:C-defined}
\end{equation}
$C$ is the entropy of the causal states, quantifying the structure present in
the observed spikes.  This is distinct from the entropy of the spikes
themselves, which quantifies not their structure but their randomness (and is
approximated by the other two terms).  Intuitively, $C$ is the (time-averaged)
amount of information about the past of the system which is relevant to
predicting its future.  For example, consider again the IID 40 Hz Bernoulli
process of Figure \ref{fig:1} A.  With $p=0.04$, this has an entropy of $0.24$
bits/msec, but because it can be described by a single state, the complexity is
zero.  (That state emits either a ``0'' or a ``1'', with respective
probabilities $0.96$ and $0.04$, but either way the state transitions back to
itself.)  In contrast, adding a 5 ms refractory period to the process means six
states are needed to describe the spike trains (Figure \ref{fig:1} B).  The new
structure of the refractory period is quantified by the higher complexity,
$C=1.05$ bits.

The second and third terms in Eq.\ \ref{eqn:final-decomp-of-average-complexity}
both describe randomness, but of distinct kinds.  The second term, the {\em
  internal entropy rate} $J$, quantifies the randomness in the state
transitions; it is the entropy of the next state given the current state.
\begin{equation}
J = H[S_{t+1}|S_t] = -\Expect{\log{\Prob(S_{t+1}|S_t)}}
\end{equation}  
This is the average number of bits per time-step needed to describe the
sequence of states the process moved through (beyond those given by $C$).  The
last term in Eq.\ \ref{eqn:final-decomp-of-average-complexity} accounts for any
residual randomness in the spikes which is not captured by the state
transitions.
\begin{equation}
R = H[X_{t+1}|S_t,S_{t+1}] = -\Expect{\log{\Prob(X_{t+1}|S_t,S_{t+1})}}
\end{equation}
For long trains, the entropy of the spikes, $H[X_1^n]$, is approximately the sum
of these two terms, $H[X_1^n] \approx n\left(J +R \right)$.  Computationally,
$C$ represents the fixed generating structure of the process, which needs to be
described once, at the beginning of the time series, and $n(J+R)$ represents
the growing list of details which pick out a particular time series from the
ensemble which could be generated; this needs, on average, $J+R$ extra bits per
time-step.  (Cf.\ the ``sophistication'' of \textcite{Gacs-Tromp-Vitanyi}.)

Consider again the 40 Hz Bernoulli process.  As there is only one state, the
process always stays in that state.  Thus the entropy of the next state $J =
0$.  However, the state sequence yields no information about the emitted
symbols (the process is IID), so the residual randomness $R=0.24$ bits/msec ---
as it must be, since the total entropy rate is $0.24$ bits/msec.  In contrast,
the states of the 5 msec refractory process are informative about the process's
future.  The internal entropy rate $J = 0.20$ bits/msec and the residual
randomness $R=0$.  All of the randomness is in the state transitions,
because they uniquely define the output spike train.  The randomness in the
state transition is confined to state $A$, where the process ``decides''
whether it will stay in $A$, emitting no spike, or emit a spike and go to $B$.
The decision needs, or gives, $0.24$ bits of information.  The transitions from
$B$ through $F$ and back to $A$ are fixed and contribute 0 bits, reducing the
expected $J$.

The important point is that the structure present in the refractory period
makes the spike train less random, lowering its entropy.  Averaged over time,
the mean firing rate of the process is $p=0.0333$.  Were the spikes IID, the
entropy rate would be $0.21$ bits/msec, but in fact $J+R=0.20$ bits/msec.
This is because a minimal description of a long sequence $X_{t_1} ... X_{t_N} =
X_{t_1}^{t_N}$, the generating process only needs to be described {\em once}
($C$), while the internal entropy rate and randomness need to be updated at
each time step ($n(J+R)$).  Simply put, a complex, structured spike train can
be exactly described in fewer bits than one which is entirely random.  The CSM
lets us calculate this reduction in algorithmic information, and quantify the
structure by means of the complexity.

\subsection{Time-Varying Complexity and Entropies}
\label{sec:time-varying-compl}

The complexity and entropy are ensemble-averaged quantities.  In the previous
section the ensemble was the entire time series, and the averaged complexity
and entropies were analogous to a mean firing rate.  The time-varying
complexity and entropies are also of interest, for example their variation
after stimuli.  A peri-stimulus time histogram (PSTH) shows how the firing
probability varies with time; the same idea works for the complexity and
entropy.

Since the states form a Markov chain, and any one spike train stays within a
single ergodic component, we can invoke the ergodic theorem
\citep{Gray-ergodic-properties}, and (almost surely) assert that
\begin{eqnarray}
\sum_{S_t,S_{t+1}} \Prob(S_t,S_{t+1},X_{t+1}) f(S_t,S_{t+1},X_{t+1}) &=& 
\lim_{N\rightarrow\infty}{\frac{1}{N} \sum_{t=1}^N f(S_t,S_{t+1},X_{t+1})} 
\nonumber\\
  &=& \lim_{N\rightarrow\infty}{\TimeAvg{f(S_t,S_{t+1},X_{t+1})}_N}
\end{eqnarray}
for arbitrary integrable functions $f(S_t,S_{t+1},X_{t+1})$.

In the case of the mean firing rate, the function to time-average is
$l(t)\equiv X_{t+1}$.  For the time averaged-complexity, internal entropy and
residual randomness, the functions (respectively $c$, $j$ and $r$) are
\begin{eqnarray}
  c(t) &=& -\log{\Prob(S_t)} \nonumber\\
  j(t) &=& -\log{\Prob(S_{t+1}|S_t)} \nonumber\\
  r(t) &=& -\log{\Prob(X_{t+1}|S_t,S_{t+1})}
\label{eqn:time-varying-defns}
\end{eqnarray}
and time-varying entropy $h(t)=j(t)+r(t)$.

The PSTH averages over an ensemble of stimulus presentations, rather than time:
\begin{eqnarray}
\lambda_{PSTH}(t) = \frac{1}{M} \sum_{i=1}^M{l_i(t)} = \frac{1}{M}
\sum_{i=1}^M{X_{t+1,i}}
\end{eqnarray}
with $M$ being number of stimulus presentations, and $t$ re-set to zero at each
presentation.  Analogously, the ``PSTH'' of the complexity is
\begin{eqnarray}
C_{PSTH}(t) = \frac{1}{M}\sum_{i=1}^M{c_i(t)} =
\frac{1}{M}\sum_{i=1}^M{-\log{\Prob(S_{t,i})}}
\end{eqnarray}
For the entropies, replace $c$ with $j$, $r$ or $h$ as appropriate.  Similar
calculations can be made with any well-defined ensemble of reference times, not
just stimulus presentations; we will also calculate $c$ and the entropies as
functions of the time since the latest spike.

We can estimate the error these time-dependent quantities as the standard error
of the mean as a function of time, $SE_t = s_t/\sqrt{M}$ where $s_t$ is the
sample standard deviation in each time bin $t$ and $M$ is the number of trials.
The probabilities appearing in the definitions of $c(t)$, $j(t)$, $r(t)$ also
have some estimation errors, either because of sampling noise or, more
interestingly, because the ensemble is being distorted by outside influences.
The latter creates a gap between their averages (over time or stimuli) and what
the CSM predicts for those averages.  In the next section, we explain how to
use this to measure the influence of external drivers.

\subsection{The Influence of External Forces}
\label{subsec:influence-of-externals}

If we know that $S_t =s$, the CSM predicts that firing probability is
$\lambda(t) = \Prob(X_{t+1} = 1|S_t=s)$.  By means of the CSM's recursive
filtering property (Appendix \ref{app:filtering}), once a transient regime has
passed, the state is always known with certainty.  Thereafter, the CSM predicts
what the firing probability should be at all times, incorporating the effects
of the spike train's history.  As we show in the next section, these
predictions give good matches to the actual response function in simulations
where the spiking probability depends only on the spike history.  But real
neurons' spiking rates generally also depend on external processes, e.g.,
stimuli.  As currently formulated, the CSM is (or, rather, converges on) the
optimal predictor of the future of the process given its {\em own} past.  Such
an ``output only'' model does not represent the (possible) effects of other
processes, and so ignores external covariates and stimuli.  Presently,
determining the precise form of spike trains' responses to external forces is
best left to parametric models.

However, we can use output-only CSMs to learn something about the computation:
the PSTH-calculated entropy rate $H_{PSTH}(t) = J_{PSTH}(t) + R_{PSTH}(t)$
quantifies the extent to which external processes drive the neuron.  (The PSTH
subscript is henceforce supressed.)  Suppose we know the true firing
probability $\lambda_{true}(t)$.  At each time step, the CSM predicts the
firing probability $\lambda_{CSM}(t)$.  If
$\lambda_{CSM}(t)=\lambda_{true}(t)$, then the CSM correctly describes the
spiking and the PSTH entropy rate is
\begin{equation}
H_{CSM}(t) = -\lambda_{CSM}(t)\log{\left[\lambda_{CSM}(t)\right]} -
(1-\lambda_{CSM}(t))\log{\left[1-\lambda_{CSM}\right(t)]}
\end{equation}
However, if $\lambda_{CSM}(t) \ne \lambda_{true}(t)$, then the CSM
mis-describes the spiking, because it neglects the influence of external
processes.  Simply put, the CSM has no way of knowing when the stimuli happen.
The PSTH entropy rate calculated using the CSM becomes
\begin{equation}
 H_{CSM}(t) = -\lambda_{true}(t)\log{[\lambda_{CSM}(t)]} -
(1-\lambda_{true}(t))\log{[1-\lambda_{CSM}(t)]}
\end{equation}
Solving $\lambda_{true}(t)$,
\begin{equation}
\lambda_{true}(t) = \frac{H_{CSM}(t) +\log{[1-\lambda_{CSM}(t)]}}
{\log{[1-\lambda_{CSM}(t)]}-\log{[\lambda_{CSM}(t)]}}
\end{equation}  

The discrepancy between $\lambda_{CSM}(t)$ and $\lambda_{true}(t)$ indicates
how much of the apparent randomness in the entropy rate is actually due
to external driving.  The true PSTH entropy rate $H_{true}(t)$ is
 \begin{equation}
 H_{true}(t) = -\lambda_{true}(t)\log{[\lambda_{true}(t)]} -
(1-\lambda_{true}(t))\log{[1-\lambda_{true}(t)]}
\end{equation}
The difference between $H_{CSM}(t)$ and $H_{true}(t)$ quantifies, in bits, the
driving by external forces as a function of the time since stimulus
presentation.
\begin{eqnarray}
\Delta H &=& H_{CSM}(t) - H_{true}(t) \nonumber\\ &=&
\lambda_{true}(t)\log{\left[\frac{\lambda_{true}(t)}{\lambda_{CSM}(t)}\right]} +
(1-\lambda_{true}(t))\log{\left[\frac{1-\lambda_{true}(t)}{1-\lambda_{CSM}(t)}\right]}
\end{eqnarray}
This {\em stimulus-driven entropy} $\Delta H$ is the relative entropy or
Kullback-Leibler divergence $D\left(X_{true}\|X_{CSM}\right)$ between the true
distribution of symbol emissions and that predicted by the CSM.
Information-theoretically, this relative entropy is the error in our prediction
of the next state due to assuming the neuron is running autonomously when it's
actually externally driven.  Since every state corresponds to a distinct
distribution over future behavior, this is our error in predicting the future
due to ignorance of the stimulus.\footnote{\normalsize 
Cf.\ the {\em informational
    coherence} introduced by \textcite{Info-Coh-for-NIPS} to measure of
  information-sharing between neurons, by quantifying the error in predicting
  the distribution of the future of one neuron due to ignoring its coupling
  with another.}

\section{Results}

We now present a few examples.  (All of them use a time-step of 1 millisecond.)
We begin with idealized model neurons to illustrate our technique.  We recover
CSMs for the model neurons using only the simulated spike trains as input to
our algorithms.  From the CSM we calculate the complexity, entropies, and, when
appropriate, stimulus-driven entropy (Kullback-Leibler divergence between the
true and CSM predicted firing probabilities) of each model neuron.  We then
analyze spikes recorded {\em in vivo} from a neuron in layer II/III of rat SI
(barrel) cortex.  We use spike trains recorded both with and without external
stimulation of the rat's whiskers.  See \textcite{Andermann} for experimental
details.

\subsection{Model neuron with a ``soft'' refractory period and bursting}
\label{sec:model-spontaneous-neuron}

\begin{figure}[p]
\includegraphics{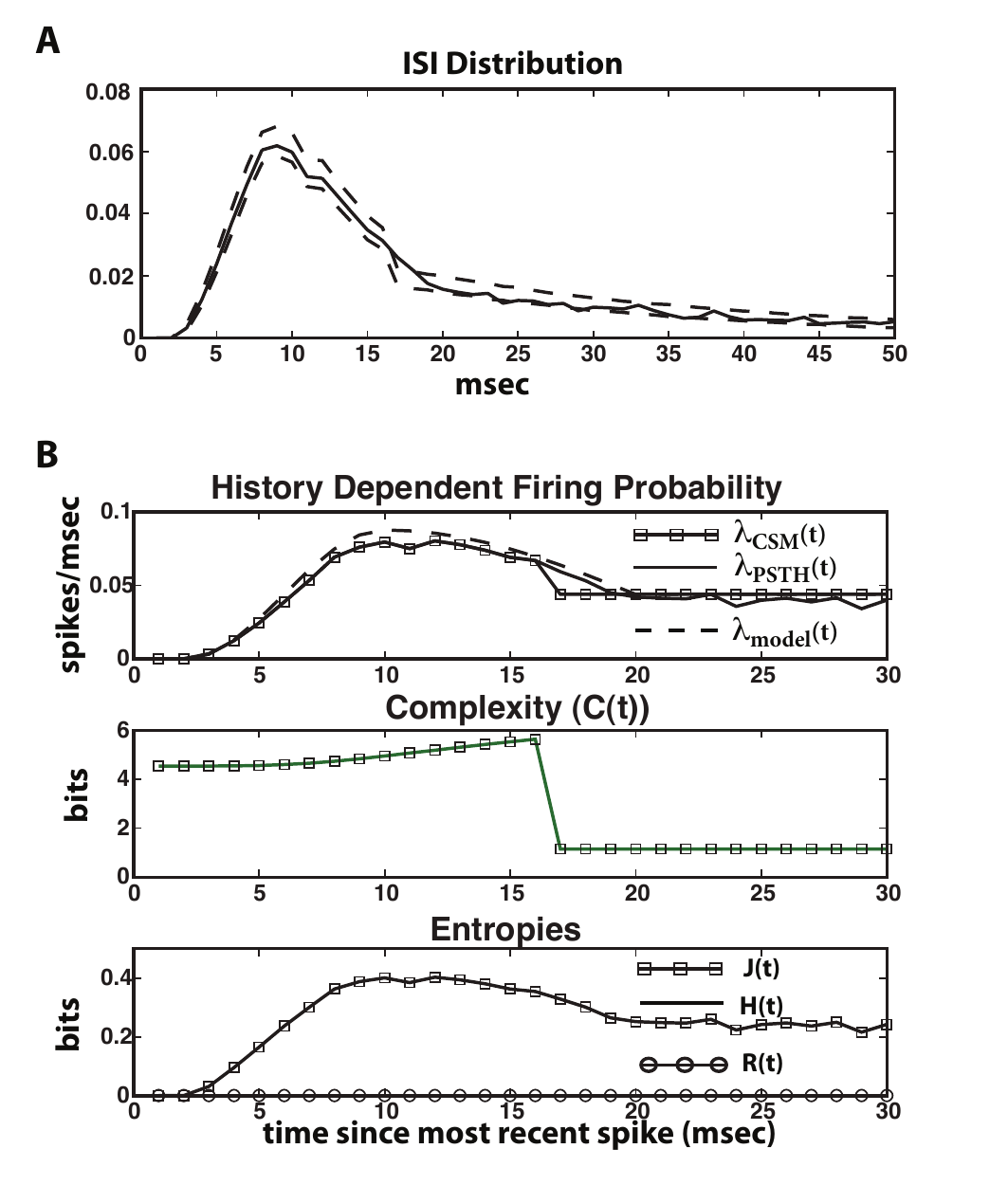}
\caption {``Soft'' refractory and bursting model ISI distribution and time
  dependent firing probability, complexity and entropies.  (A) ISI distribution
  and 99\% confidence bounds bootstrapped from the CSM.  (B) First panel:
  Firing probability as a function of time since the most recent spike.  Line
  with squares = firing probability predicted by CSM.  Solid line = firing
  probability deduced from PSTH. Dashed line = model firing rate used to
  generate spikes.  Second panel: Complexity as a function of time since most
  recent spike.  Third panel: Entropies as a function of time since most recent
  spike.  Squares = internal entropy rate, circles = residual randomness, solid
  line = entropy rate.  (overlaps squares)}
\label{fig:3}
\end{figure}

We begin with a refractory, bursting model neuron, whose spiking rate depends
only on the time since the last spike.  The baseline rate is 40 Hz.  Every
spike is followed by a 2 msec ``hard'' refractory period, during which spikes
never occur.  The spiking rate then rebounds to twice its baseline, to which it
slowly decays.  (See dashed line in the first panel of Figure \ref{fig:3} B.)
This history dependence mimics that of a bursting neuron, and is, intuitively,
more complex than the simple refractory period of the model in Figure
\ref{fig:1}.

\begin{figure}[p]
\includegraphics{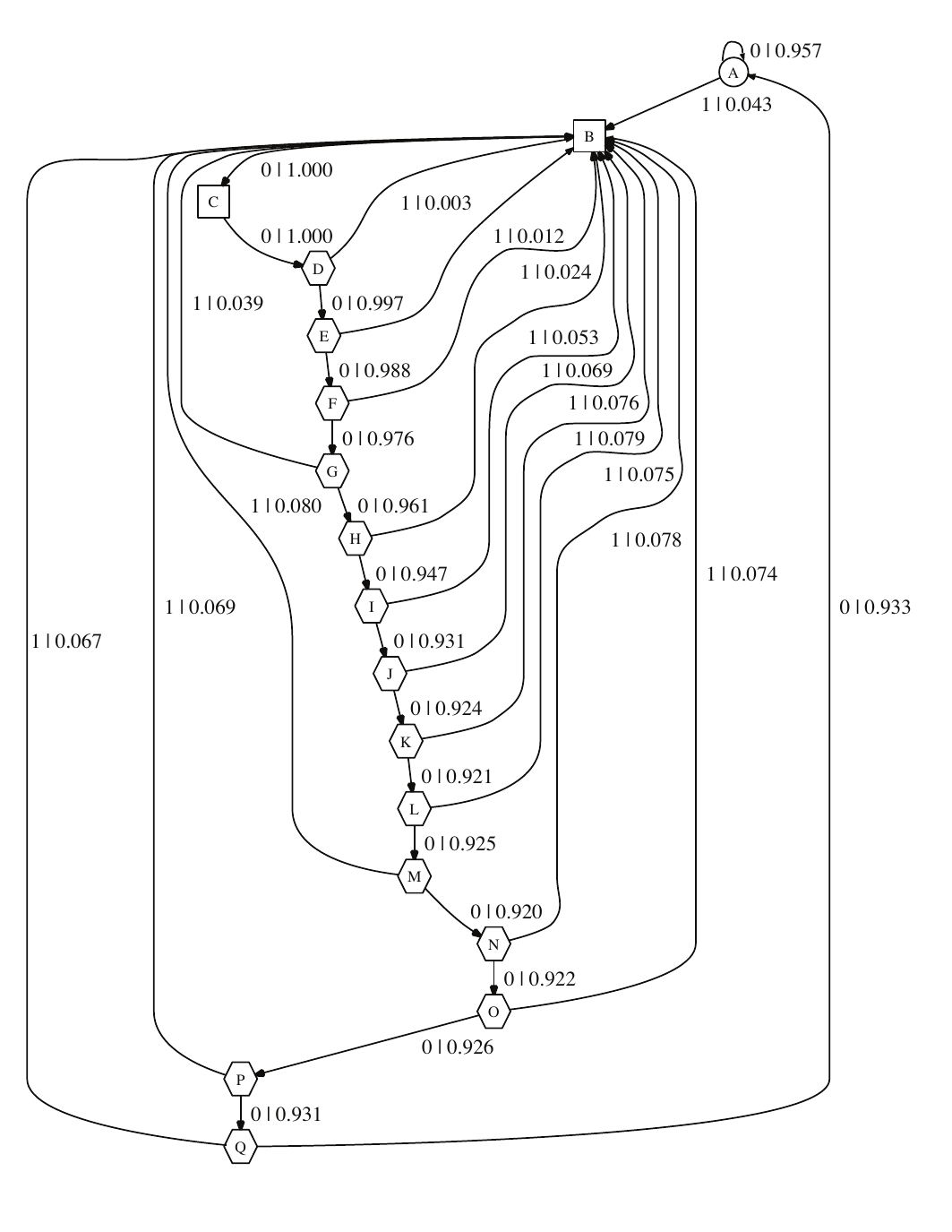}
\caption{CSM reconstructed from a 200 sec simulated spike train with a ``soft''
  refractory/bursting structure.  $C=3.16$, $J=0.25$, $R=0$.  State $A$
  (circle) is the baseline 40 Hz spiking state.  Upon emitting a spike the
  transition is to state $B$.  States $B$ and $C$ (squares) are ``hard''
  refractory states from which no spike may be emitted. States $D$ through $Q$
  (hexagons) compromise a refractory/bursting chain from which if a spike is
  emitted the transition is back to state $B$.  Upon exiting the chain the CSM
  returns to the baseline state $A$.}
\label{fig:2}
\end{figure}

Figure \ref{fig:2} shows the 17-state CSM reconstructed from a 200 second
spike train (at 1 msec resolution)
generated by this model.  It has a complexity of $C=3.16$ bits (higher than
that of the model in Figure \ref{fig:1}, as anticipated), an internal entropy
rate of $J=0.25$ bits/msec and a residual randomness of $R=0$ bits/msec.  The
CSM was obtained with $\Lmax=17$ (selected by BIC).  Figure \ref{fig:3} A shows
how the 99\% ISI bounds bootstrapped from the CSM enclose the empirical ISI
distribution, with the exception of one short segment.

The CSM is easily interpreted.  State $A$ is the baseline state.  When it emits
a spike, the CSM moves to state $B$.  There are then two deterministic
transitions, to $C$ and then $D$, which never emit spikes; this is the hard 2
msec refractory period.  Once in $D$ it is possible to spike again, and if that
happens, the transition is back to state $B$.  However, if no spike is emitted,
the transition is to state $E$.  This is repeated, with varying firing
probabilities, as states $E$ through $Q$ are traversed.  Eventually, the
process returns to $A$ and so to baseline.

Figure \ref{fig:3} B plots the firing rate, complexity, and internal entropies
as functions of the time since the last spike {\em conditional on no subsequent
  spike emission}.  This lets us compare the firing rate predicted by the CSM
(solid line squares) to the specification of the model which generated the
spike train (dashed line) and a PSTH calculated by triggering on the last spike
(solid line).  Except at 16 and 17 msec post spike, the CSM-predicted firing
rate agrees with both the generating model and the PSTH.  The discrepancy
arises because the CSM only discerns the structure in the data, and most of the
ISIs are shorter than 16 msec.  There is much closer agreement between the CSM
and the PSTH if firing rates are plotted as a function of time since a spike
without conditioning on no subsequent spike emission (not shown).

The second and third panels of Figure \ref{fig:3} plot the time-dependent
complexity and entropies.  The complexity is much higher after the emission of
a spike than during baseline, because the states traversed (B-Q) are less
probable, and represent the additional structures of refractoriness and
bursting.  The time-dependent entropies (third panel) show that just after a
spike, the refractory period imposes temporary determinism on the spike train,
but burstiness increases the randomness before the dynamics return to the
baseline state.

\subsection{Model neuron under periodic stimulation}
\label{sec:model-stimulated-neuron}

\begin{figure}[p]
\includegraphics{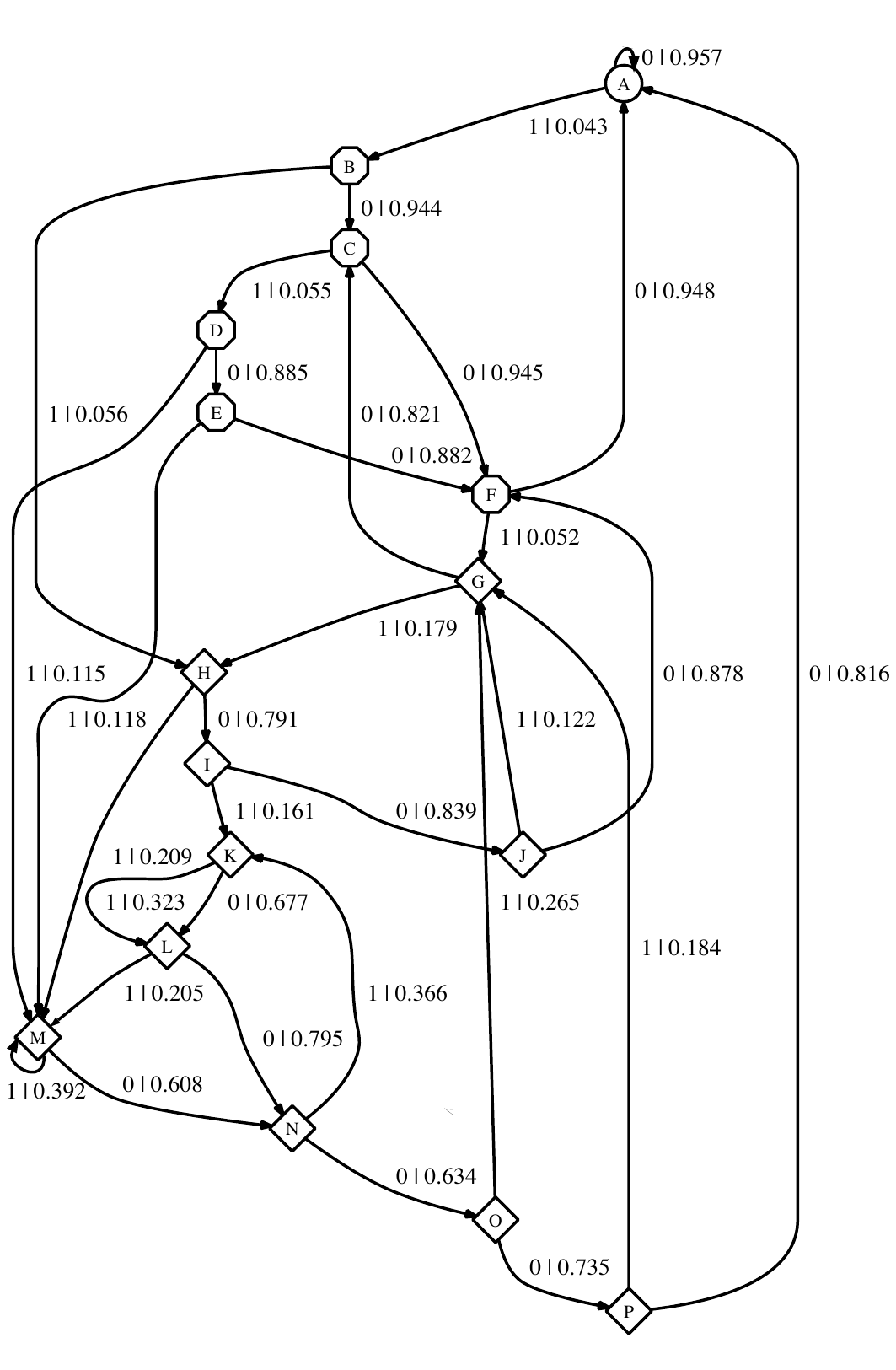}
\caption {16-state CSM reconstructed from 200 sec of simulation of
  periodically-stimulated spiking.  $C=0.89$, $J=0.27$, $R=0.0007$.  State $A$
  is the baseline state.  States $B$ through $F$ (octagons) are ``decision''
  states in which the CSM evaluates whether a spike indicates a stimulus or was
  spontaneous.  Two spikes within 3 msec cause the CSM to transition to states
  $G$ through $P$, which represent the structure imposed by the stimulus. If no
  spikes are emitted within 5 (often fewer) sequential msec, the CSM goes back
  to the baseline state $A$.}
\label{fig:4}
\end{figure}

\begin{figure}[p]
\includegraphics{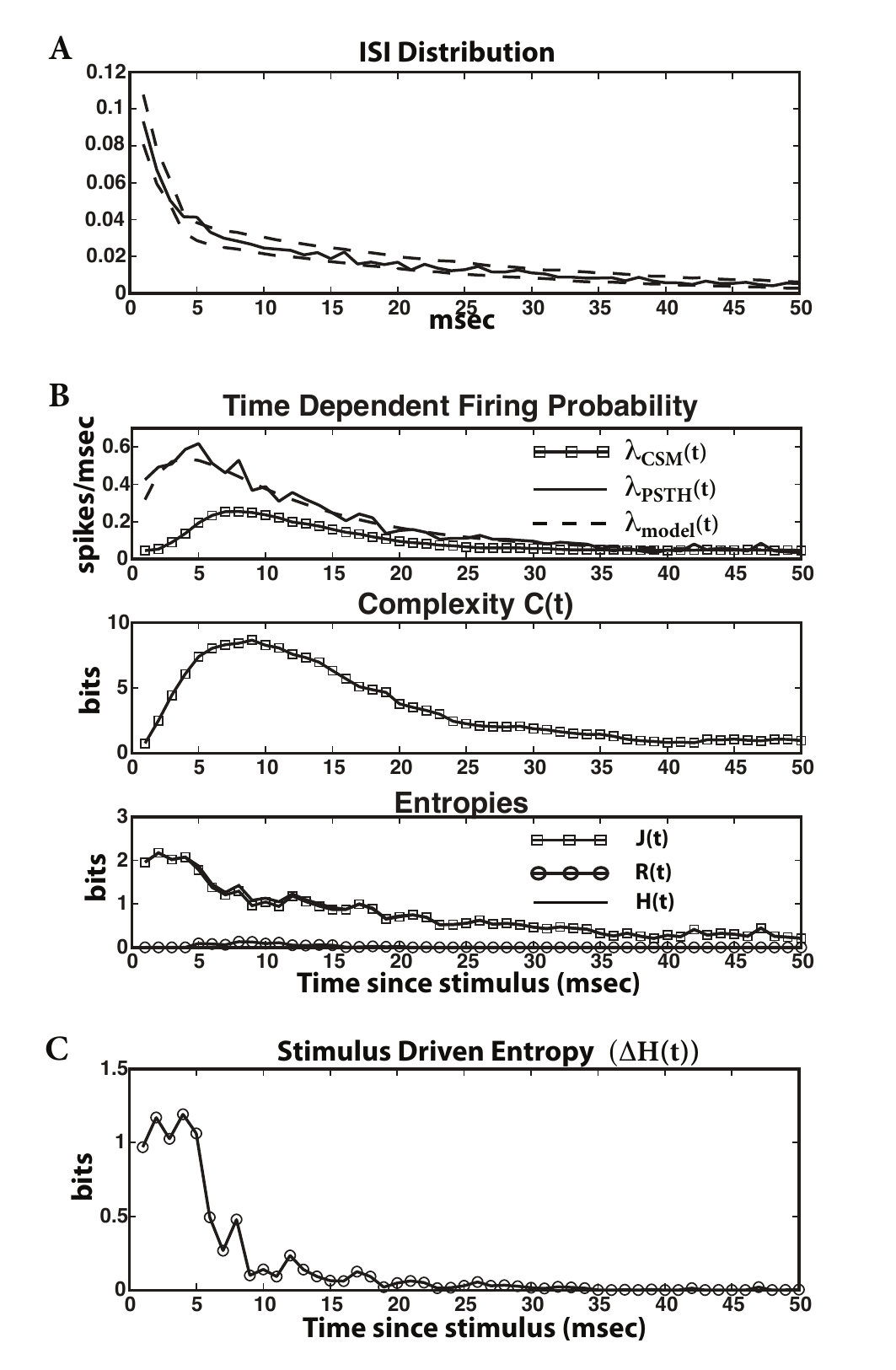}
\caption{Stimulus model ISI distribution and time-dependent complexity and
  entropies.  (A) ISI distribution and 99\% confidence bounds.  (B) First
  panel: Firing probability as a function of time since stimulus presentation.
  Second panel: Time dependent complexity.  Third panel: time-dependent
  entropies.  (C) The stimulus-driven entropy is $> 1$ bit, indicating strong
  external drive. See text for discussion.}
\label{fig:5}
\end{figure}

Figure \ref{fig:4} shows the CSM for a periodically-stimulated model neuron .
This CSM was reconstructed from 200 seconds of spikes with a baseline firing
rate of 40 Hz ($p=0.04$).  Each second, the firing rate rose over the course of
5 msec to $p=0.54$ spikes/msec, falling slowly back to baseline over the next
50 msec.  This mimics the periodic presentation of a strong external stimulus.
(The exact inhomogeneous firing rate used was $\lambda(t) =
0.93[e^{-t/10}-e^{-t/2}] + 0.04$ with $t$ in msec.  See Figure \ref{fig:5} B,
first panel, dashed line.)  In this model, the firing rate does not directly
depend on the spike train's history, but there is a sort of history dependence
in the stimulus time-course, and this is what CSSR discovers.

BIC selected $\Lmax=7$, giving a 16 state CSM with $C=0.89$ bits, $J=0.27$
bits/msec and $R=0.0007$ bits/msec.  The baseline is again state $A$ and if no
spike is emitted then the process stays in $A$.  Spikes are either spontaneous
and random, or stimulus-driven.  Because the stimulus is external, it is not
immediately clear which of these two causes produced a given spike.  Thus, if a
spike is emitted, the CSM traverses states $B$ through $F$, deciding, so to
speak, whether or not the spike is due to a stimulus.  If two spikes happen
within 3 msec of each other, then the CSM decides that it is being stimulated
and goes to one of states $G$, $H$ or $M$.  States $G$ through $P$ represent
the response to the stimulus.  The CSM moves between these states until no
spike is emitted for 3 msec, when it returns to the baseline, $A$.

The ISI distribution from the CSM matches that from the model (Figure
\ref{fig:5} A).  However, because the stimulus doesn't depend on spike train's
history, the CSM makes inaccurate predictions during stimulation.  The first
panel of Figure \ref{fig:5} B plots the firing rate as a function of time since
stimulus presentation, comparing the model (dashed line) and the PSTH (solid
line) with the CSM's prediction (line with squares).  The discrepancy between
these is due to the CSM having no way of knowing that an external stimulus has
been applied until several spikes in a row have been emitted (represented, as
we just say, by states $B$--$F$)\footnote{\normalsize 
In effect, this part of the CSM
  implements Bayes's rule, balancing the increased {\em likelihood} of a spike
  after a stimulus against the low {\em a priori} probability or base-rate of
  stimulation.}.  Despite this, $c(t)$ shows that something more complex than
simple random firing is happening (second panel of Figure \ref{fig:5} B), as do
$j(t)$ and $r(t)$ (third panel).  Further, something is clearly wrong with the
entropy rate, because it should be upper-bounded by $h=1$ bit/msec (when
$p=0.5$).  The fact that $h(t)$ exceeds this bound indicates an external force,
not fully captured by the CSM, is at work.

As discussed in Methods (\S \ref{subsec:influence-of-externals}), drive from
the stimulus can be quantified with a relative entropy (Figure \ref{fig:5} C).
Stimuli are presented at $t=1$ msec, where $\Delta H(t) > 1$ bit.  It is not
until $\approx 25$ msec post-stimulus that $\Delta H(t) \approx 0$ and the CSM
once again correctly describes the internal entropy rate.  Thus, as expected,
the stimulus strongly influences neuronal dynamics immediately after its
presentation.  The true internal entropy rate $H_{true}(t)$ is slightly less
than 1 bit/msec shortly after stimulation, when the true spiking rate has a
maximum of $p_{max}=0.54$.  The fact that the CSM gives an inaccurate value for
$J$ actually lets us find the number of bits of information gain supplied by
the stimulus, e.g., $\Delta H > 1$ bit immediately after the stimulus is
presented.

\subsection{Spontaneously Spiking Barrel Cortex Neuron}

\begin{figure}[p]
\includegraphics{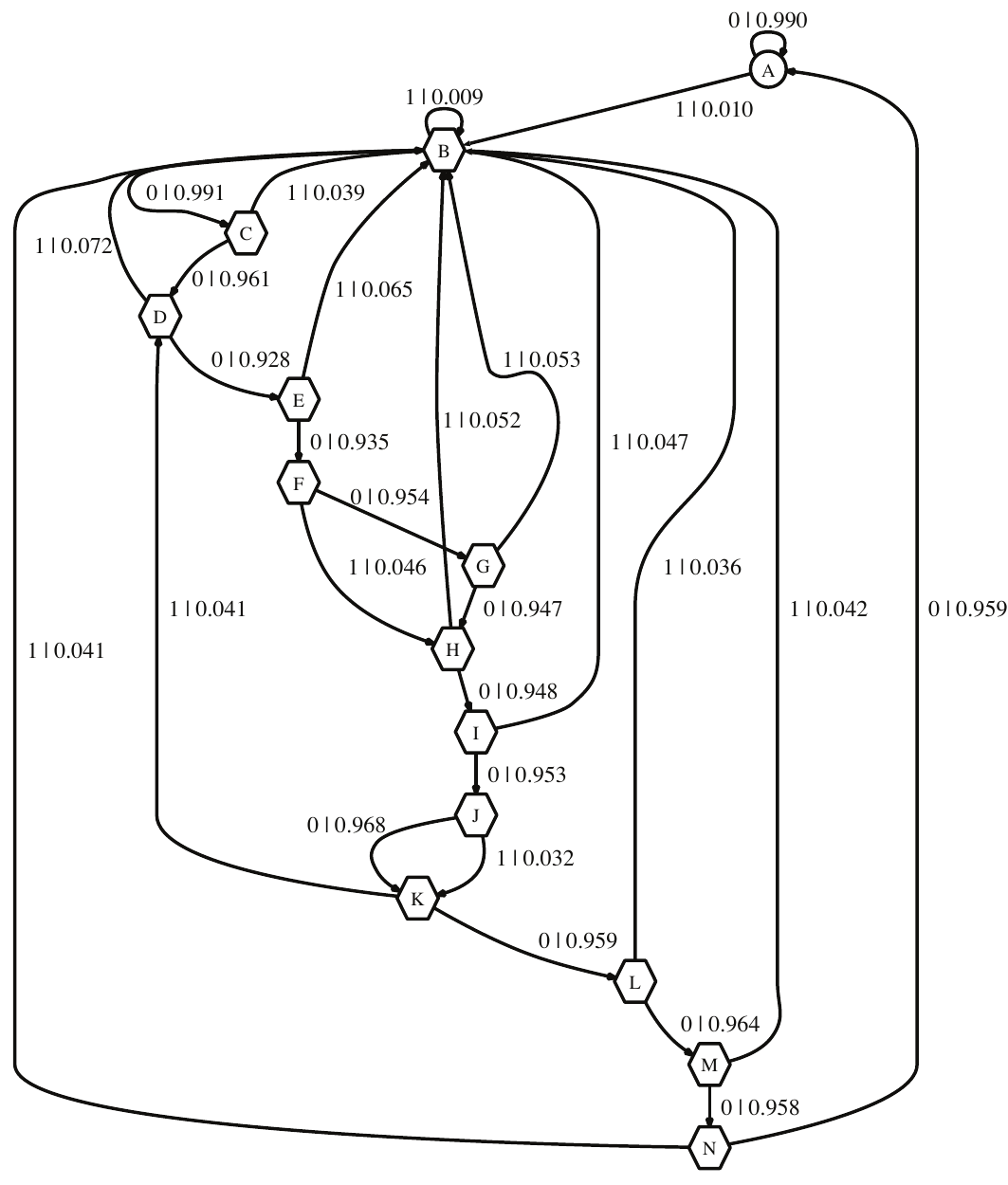}
\caption{14-state CSM reconstructed from 90 sec of spiking recorded from a
  spontaneously spiking (no stimulus) neuron located in layer II/III of rat
  barrel cortex.  $C= 1.02$, $J= 0.10$, $R=0.005$. State $A$ (circle) is
  baseline 10 Hz spiking.  States $B$ through $N$ comprise a
  refractory/bursting chain similar to, but with a somewhat more intricate
  structure than, that of the model neuron in Figure \ref{fig:2} }
\label{fig:6}
\end{figure}

We reconstructed a CSM from 90 seconds of spontaneous (no vibrissa deflection)
spiking recorded from a layer II/III FSU barrel cortex neuron.  CSSR, using
$\Lmax=21$, discovered a CSM with 315 states, a complexity of $C=1.78$ bits,
and internal entropy rate of $J=0.013$ bits/msec.  After state culling (\S
\ref{section:condense}), the reduced CSM, plotted in Figure \ref{fig:6}, has 14
states, $C=1.02$, $J=0.10$ bits/msec, and residual randomness of $R=0.005$
bits/msec.  We focus on the reduced CSM from this point onwards.

This CSM resembles that of the spontaneously-firing model neuron of \S
\ref{sec:model-spontaneous-neuron} and Fig. \ref{fig:2}.  The complexity and
entropies are lower than those of our model neuron because the mean spike rate
is much lower, and so simple descriptions suffice most of the time. (Barrel
cortex neurons exhibit notoriously low spike rates, especially during
anesthesia.)  There is a baseline state $A$ which emits a spike with
probability $p=0.01$, i.e., 10 Hz.  When a spike is emitted, the CSM moves to
state $B$ and then on through the chain of states $C$ through $N$, return to
$A$ if no spike is subsequently emitted.  However, the CSM can emit a second or
even third spike after the first, and indeed this neuron displays spike
doublets and triplets.  In general, emitting a spike moves the CSM to $B$, with
some exceptions that show the structure to be more intricate than the model
neuron's.

\begin{figure}[p]
\includegraphics{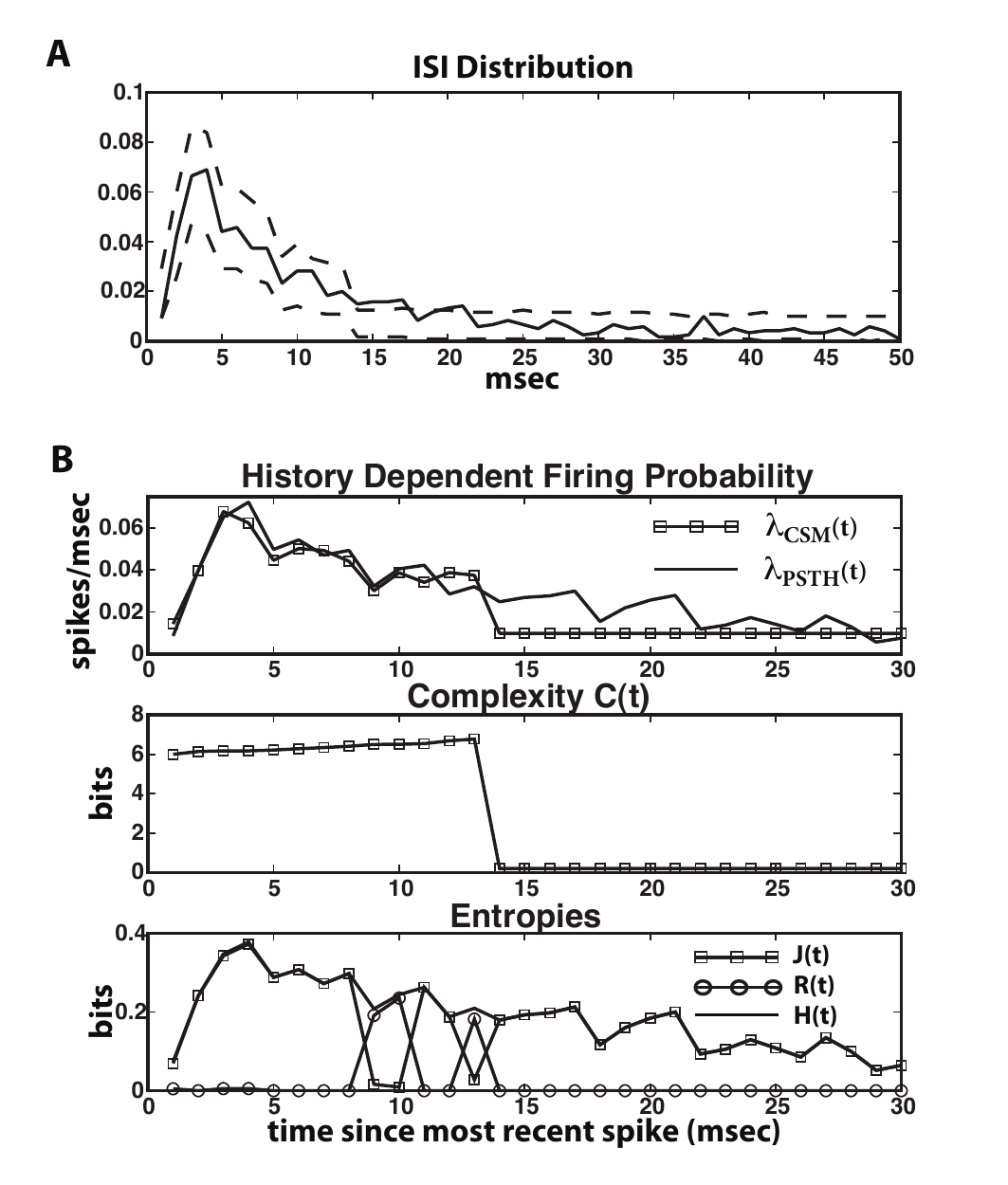}
  \caption{Spontaneously spiking barrel cortex neuron.  (A) ISI distribution
    and 99\% bootstrapped confidence bounds.  (B) First panel: Time dependent
    firing probability as a function of time since most recent spike.  See text
    for explanation of discrepancy between CSM and PSTH spike probabilities.
    Second Panel: Complexity as a function of time since most recent spike.
    Third Panel: Entropy rates as a function of time since most recent spike.}
\label{fig:7}
\end{figure}

\begin{figure}[p]
\includegraphics{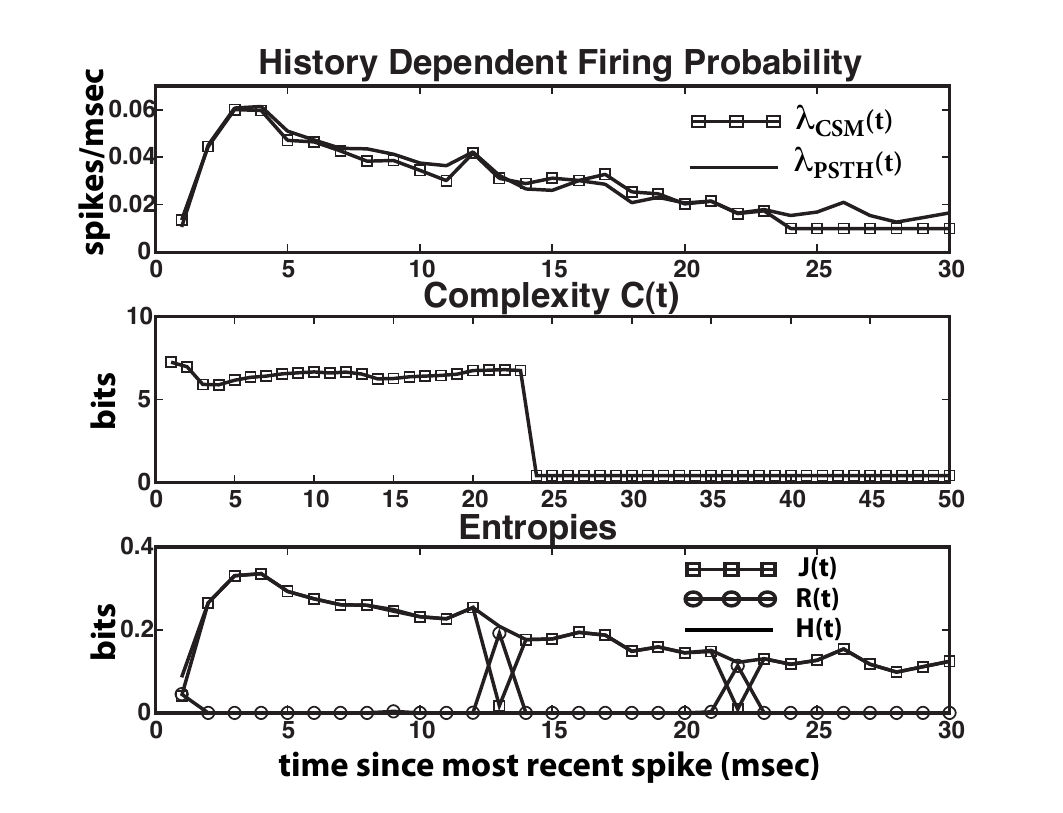}
\caption{Firing probability, complexity and entropies of the 
   stimulated barrel cortex
  neuron as a function of time since the most recent spike.}
\label{fig:10}
\end{figure}

Figure \ref{fig:7} A shows the CSM's 99\% confidence bounds almost completely
enclosing the empirical ISI distribution.  The first panel of Figure
\ref{fig:7} B plots the history-dependent firing probability predicted by the
CSM as a function of the time since the latest spike, according to both the
PSTH and the CSM's prediction.  They are highly similar in the first 13 msec
post-spike, indicating that the CSM gets the spiking statistics right in this
epoch.  The CSM and PSTH the diverge after this, for two reasons.  First, as
with the model neuron, there are few ISIs of this length.  Most of the ISIs are
either shorter, due to the nueron's burstiness, or much longer, due to the low
baseline firing rate.  Secondly, 90 seconds is not very much data.  We show in
Figure \ref{fig:10} that a CSM reconstructed from a longer spike train does
capture all of the structure.  We present the results of this shorter spike
train to emphasize that, as a non-parametric method, CSSR only uncovers the
statistical structure {\em in the data}, no more, no less.

Finally, the second and third panels of Figure \ref{fig:6} B show,
respectively, the complexity and entropies as functions of the time since the
latest spike.  As with the model of \S \ref{sec:model-spontaneous-neuron},
the structure in the process occurs after spiking, during the refractory and
bursting periods.  This is when the complexity is largest, and also when the
entropies vary most.

\subsection{Periodically Stimulated Barrel Cortex Neuron}

\begin{figure}[p]
\includegraphics{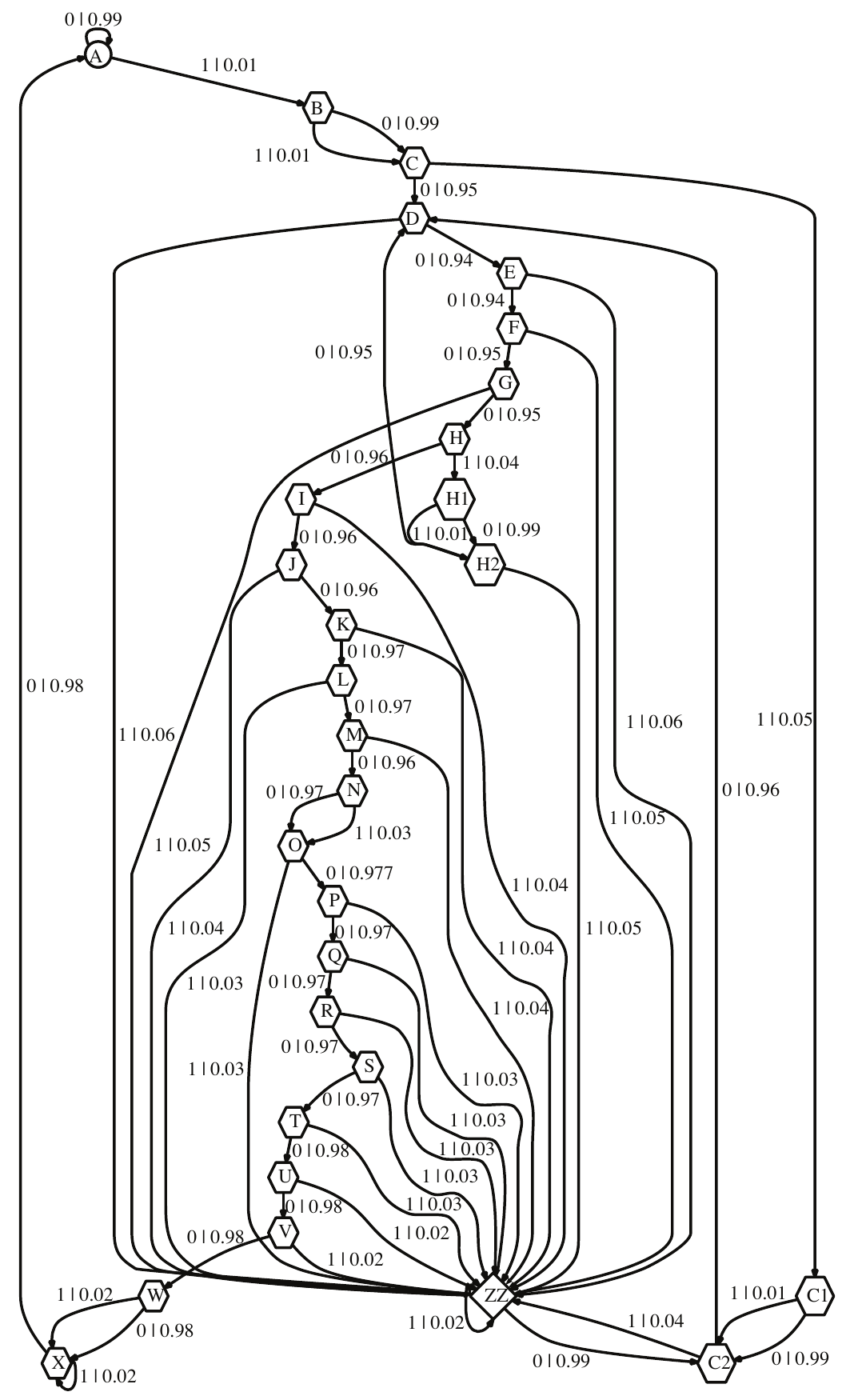}
\caption{29-state CSM reconstructed from 335 seconds of spikes recorded from a
  layer II/III barrel cortex neuron undergoing periodic (125 msec
  inter-stimulus interval) stimulation via vibrissa deflection.  $C=1.97$,
  $J=0.11$, $R=0.004$.  Most of the states are devoted to refractory/bursting
  behavior, however states ``C1'', ``C2'' and ``ZZ'' represent the structure
  imposed by the external stimulus.  See text for discussion.}
\label{fig:8}
\end{figure}

We reconstructed CSMs from 335 seconds of spike trains taken from the same
neuron used above, but recorded while it was being periodically stimulated by
vibrissa deflection.  BIC selected $\Lmax=25$, giving the 29-state CSM shown in
Figure \ref{fig:8}.  (Before state culling, the original CSM had 1916 states,
$C=2.55$ and $J=0.11$.)  The reduced CSM has a complexity of $C=1.97$ bits, an
internal entropy rate of $J=0.10$ bits/msec, and a residual randomness of
$R=0.005$ bits/msec.  Note that $C$ is higher when the neuron is being
stimulated as opposed to when it is spontaneously firing, indicating more
structure in the spike train.

While at first the CSM may seem to only represents history-dependent
refractoriness and bursting, ignoring the external stimulus, this is not quite
true.  Once again, there is a baseline state $A$, and most of the other states
($B$--$X$) comprise a refractory/bursting chain, like this neuron has during
spontaneous firing.  However, the transition upon $A$ emitting a spike is not
back to $B$ and then down the chain again, but to either state $C_1$, and
subsequently $C_2$, or more often to state $ZZ$.  These three states represent
the structure induced by the external stimulus, as we saw with the model
stimulated neuron of \S \ref{sec:model-stimulated-neuron} and Figure
\ref{fig:4}.  (The state $ZZ$ is comparable to the state $M$ of the model
stimulated neuron: both loop back to themselves if they emit a spike.)  Three
states are enough because, in this experiment, barrel cortex neurons spike
extremely sparsely, $0.1$--$0.2$ spikes per stimulus presentation.

\begin{figure}[p]
\includegraphics{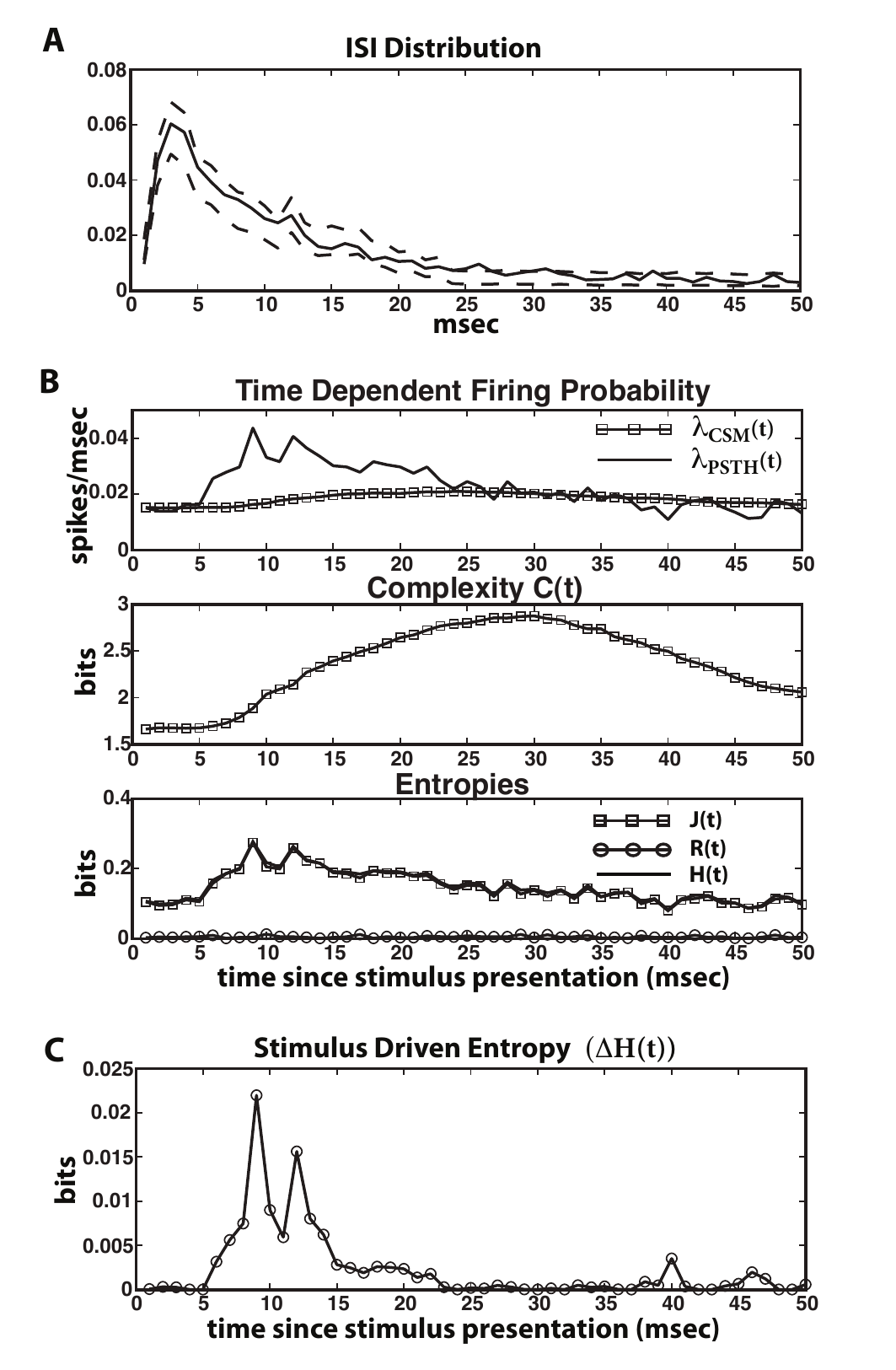}
\caption{Stimulated barrel cortex neuron ISI distribution and time-dependent
  complexity and entropies.  (A) ISI distribution and 99\% confidence bounds.
  (B) First panel: Firing probability as a function of time since stimulus
  presentation.  Second panel: Time-dependent complexity.  Third panel:
  time-dependent entropies.  (C) The stimulus driven entropy (maximum of $0.02$
  bits/msec) is low because the number of spikes per stimulus
  ($\approx 0.1 - 0.2$) is very low and hence the stimulus does not supply much
  information.}
\label{fig:9}
\end{figure}

Figure \ref{fig:9} A plots the ISI distribution, nicely enclosed by the
bootstrapped confidence bounds.  Figure \ref{fig:9} B shows the firing rate,
complexity and entropies as functions of the time since stimulus presentation
(averaged over all presentations).  These plots look much like those in Figure
\ref{fig:7} B.  However, there is a clear indication that something more
complex takes place after stimulation: the CSM's firing-rate predictions are
wrong.  The stimulus-driven entropy $\Delta H$ turns out to be as large as
$0.02$ bits within 5--15 msec post-stimulus.  This agrees with the known
$\approx 5$--$10$ msec stimulus propagation time between vibrissae and barrel
cortex \citep{Andermann}.  The reason that $\Delta H$ is so much smaller for
the real neuron than the stimulated model neuron of \S
\ref{sec:model-stimulated-neuron} is that the former's firing rate is much
lower.  Although the firing rate post-stimulus can be almost twice as large as
the CSM's prediction, the actual rate is still low, $\max{\lambda(t)} \approx
0.04$ spikes/msec.  Most of the time the neuron does not spike, even when
stimulated, so on average, the stimulus provides little information per
presentation.  For completeness, Figure \ref{fig:10} shows the spike
probability, complexity and entropies as functions of the time since the latest
spike.  Averaged over this ensemble, the CSM's predictions are highly accurate.

\section{Discussion}

The goal of this paper was to present methods for determining the structural
content of spike trains while making minimal {\em a priori} assumptions as to
the form which that structure takes.  We use the CSSR algorithm to build
minimal, optimally predictive hidden Markov models (CSMs) from spike trains,
Schwartz's Bayesian Information Criterion to find the optimal history length
$\Lmax$ of the CSSR algorithm, and bootstrapped confidence bounds on the ISI
distribution from the CSM to check goodness-of-fit.  We demonstrated how CSMs
can estimate a spike train's complexity, thus quantifying its structure, and
its mean algorithmic information content, quantifying the minimal computation
necessary to generate the spike train.  Finally we showed how to quantify, in
bits, the influence of external stimuli upon the spike-generating process.  We
applied these methods both to simulated spike trains, for which the resulting
CSMs agreed with intuition, and to real spike trains recorded from a layer
II/II rat barrel cortex neuron, demonstrating increased structure, as measured
by the complexity, when the neuron was being stimulated.

We are unaware of any other practical techniques for quantifying the complexity
and computational structure of a spike train as we define them.  Intuitively,
neither random (Poisson), nor highly ordered (e.g., strictly periodic, as in
\textcite{New-roles-for-the-gamma-rhythm}) spike trains should be thought of as
complex since they do not possess structure requiring a sophisticated program
to generate.  Instead, complexity lies between order and disorder
\citep{Badii-Politi}, in the non-random variation of the spikes.  Higher
complexity means a greater degree of organization in neural activity than would
be implied by random spiking.  It is the reconstruction of the CSM through CSSR
which allows us to calculate the complexity.

Our definition of complexity stands in stark contrast to other complexity
measures which assign high values to highly disordered systems.  Some of these,
such as Lempel Ziv complexity
\citep{Amigo-LZ-Nstates,Amigo-entropy-rate,Jimenez-Entropy-LZ,
  Szczepanski-LZ-2004} and context free grammar complexity \citep{Rapp-1994}
have been applied to spike trains.  However, both of these are measures of the
amount of information required to reproduce the spike train {\em exactly}, and
take on very high values for completely random sequences.  These ``complexity''
measures are therefore much more similar to total algorithmic information
content and even to the entropy rate than to our sort of complexity.

Our measure of complexity is the entropy of the distribution of causal states.
This has the desired property of being maximized for structured, rather than
ordered or disordered systems, because the causal states are defined
statistically, as equivalence classes of histories conditioned on future
events.  Other researchers have also calculated complexity measures which are
entropies of state distributions, but have defined their states differently.
\textcite{Amigo-LZ-Nstates} uses the observables (symbol strings) present in
the spike train to define a k-th order Markov process and calls each individual
length k string which appears in the spike train a state.
\textcite{Gorse-and-Taylor} similarly use single suffix symbol strings to
define the states of a Markov process.  In both cases, IID Bernoulli sequences
could exhibit up to $2^k$ states (in long enough sequences), and possess an
extremely high ``complexity''.  However, all of these states make the same
prediction for the future of the process.  The minimal representation is a
single causal state, a CSM with a complexity of zero.

It should be noted that there are also many works which model spike trains
using HMMs, but in which the hidden states represent {\em macro}-states of the
system (awake/asleep, Up/Down, etc.), and spiking rates are modeled separately
in each macro-state
\citep{Abeles-HMM,Achtman-HMM,Chen-Up-Down,Danoczy-NIPS,Jones-HMM}.  Although
the graphical representation of such HMMs may look like those of CSMs, the two
kinds of states have very different meanings. Finally, there are also state
space methods which model the dynamical state of the system as a continuous
hidden variable, the most well known of which is the linear Gaussian model with
Kalman filtering.  These have been extensively applied to neural encoding and
decoding problems
\citep{Eden-Kalman,Smith-state,Srinivasan-Kalman}. Interestingly, for a
univariate Gaussian ARMA model in state-space form, the Kalman filter's
one-step-ahead prediction and mean-squared prediction error are, jointly,
minimal-sufficient for next-step prediction, and since they can be updated
recursively they in fact constitute the minimal sufficient statistic, and hence
the causal state in this special case.

Neurons are driven by their afferent synapses.  Although as discussed in
Appendix \ref{sec:CSTs-and-PSRs}, there is a parallel ``transducer'' formalism
for generating CSMs which take external influences into account, this is not yet
computationally implemented, and our current approach reconstructs CSMs only
from the spike train.  Since the history of the neuron under study is typically
connected with the history of the network in which it is located, this CSM
will, in general, reflect more than a neuron's internal biophysical properties.
Nonetheless, in both our model neurons and in the real barrel cortex neuron,
states not interpretable as simple refractoriness or bursting appeared when a
stimulus was present, proving we can detect stimulus-driven complexity.
Further, we showed that the CSM can be used to determine the extent (in bits)
to which a neuron is driven by external stimuli.

The methods presented here complement more established modes of spike-train
analysis, which have different goals.  Parametric methods, such as PSTHs or
maximum likelihood estimation
\citep{Brown-Kass-Mitra-multiple-spike-train-data,Truccolo-GLM} generally focus
on determining a neuron's firing rate (mean, instantaneous or
history-dependent), and on how known external covariates modulate that rate.
They have the advantage of requiring less data than non-parametric methods such
as CSSR, but the disadvantage, for our purposes, of imposing the structure of
the model at the outset.  When the experimenter wants to know how a neuron
encodes a particular aspect of a covariate, e.g., how neurons in the sensory
periphery or primary sensory cortices encode stimuli, parametric methods have
proved highly illuminating.  However, in many cases the identity or even
existence of relevant external covariates is uncertain.  For example, one could
envision using CSMs to analyze recordings in pre-frontal cortex during
different cognitive tasks, or to perhaps compare spiking structure during
different attentional states.  In both cases, the relevant external covariates
are not at all clear, but CSMs could still be used to quantify changes in
computational structure, for single neurons or for groups of them.  For neural
populations one can envision generating distributions (over the population) of
complexities and examining how these distributions change in different cortical
macro-states.  This would be entirely analagous to analyzing distributions of
firing rates or tuning curves.

In addition to calculations of the complexity, the whole array of
mutual-information analyses can be applied to CSMs, but instead of calculating
mutual information between the spikes and the covariates (which could include
other spike trains), one can calculate the mutual information between the
covariates and the {\em causal states}.  The advantage is that the causal
states represent the behavioral patterns of the spike-generating process, and
so are closer to the actual state of the system than the spikes (output
observables) are themselves.  Results on calculating the mutual information
between the causal states of different neurons (informational coherence) in a
large simulated network show that synchronous neuronal dynamics are more
effectively revealed than when calculated directly from the spikes
\citep{Info-Coh-for-NIPS}.

In closing, our methods provide a way to understand structure in spike trains,
and should be considered as complements to traditional analysis methods.  We
rigorously define structure, and show how to discover it from the data itself.
Our methods go beyond those which seek to describe the observed variation in
the spiking rates by also describing the underlying computational process (in
the form of a CSM) needed to generate that variation.  A CSM can show not only
that the spike rate has changed, but also show exactly {\em how} it has
changed.

\paragraph*{Acknowledgments}
The authors thank Mark Andermann and Christopher Moore for the use of their
data.  RH thanks Emery Brown, Anna Dreyer and Christopher Moore for valuable
discussions.  CRS thanks Anthony Brockwell, Dave Feldman, Chris
Genovese, Rob Kass and Alessandro Rinaldo for valuable discussions.

\begin{appendix}

\section{Filtering with CSMs}
\label{app:filtering}

A common difficulty with hidden Markov models is that predictions can only be
made from a knowledge of the state, which must itself be guessed at from the
time series, since it is, after all, hidden.  This creates the {\em state
  estimation} or {\em filtering} problem.  Under strong assumptions (linear
Gaussian stochastic dynamics, linearly observed through IID additive Gaussian
noise) the Kalman filter is an optimal yet tractable solution.  For non-linear
processes, however, optimal filtering essentially amounts to maintaining a
posterior distribution over the states and updating it via Bayes's rule
\citep{Ahmed-filtering}.  (This distribution is sometimes called the process's
``information state''.)

One convenient and important feature of CSMs is that this whole machinery of
filtering is unnecessary, because of their recursive-updating property.  Given
the state at time $t$, $S_t$, and the observation at time $t+1$, $X_{t+1}$, the
state at time $t+1$ is fixed, $S_{t+1} = T(S_t, X_{t+1})$ for some transition
function $T$.  Clearly, if the state is known with certainty at any time, it
will remain known.  However, the same recursive updating property also allows
us to show that the state does become certain, i.e., that after some finite
(but possibly random) time $\tau$, $\Prob(S_\tau = s|X_1^\tau)$ is either 0 or
1 for all states $s$.  For Markov chains of order $k$, clearly $\tau \leq k$;
under more general circumstances $\Prob(\tau \geq t)$ goes to zero
exponentially or faster.

Thus, after a transient period, the state is completely unambiguous.  This will
be useful to us in multiple places, including understanding the computational
structure of the process and predicting the firing rate of the neuron.  It also
leads to considerable numerical simplifications, compared to approaches which
demand conventional filtering.  Further, recursive filtering is easily applied
to a new spike train, not merely the one from which the CSM was reconstructed.
This helps in cross-validating CSMs, as discussed in the next appendix.

\section{Cross-Validation}
\label{Appendix:Cross-Validate}

It is often desirable to cross-validate a statistical model by spliting one's
data set in two, using one part (generally the larger) as a training set for
the model and the other part to validate the model by some statistical test.
In the case of CSMs it is particularly important to check the validity of the
BIC used to regularize the $\Lmax$ control-setting.

One possible test is the ISI bootstrapping of \S \ref{subsect:ISIbootstrap}.  A
second, somewhat stronger, goodness-of-fit test is based on the time rescaling
theorem of \textcite{Brown-time-rescaling}.  This test rescales the interspike
intervals as a function of the integrated history-dependent spiking rate over
the ISI:
\begin{eqnarray}
\tau_k = 1 - e^{-\int_{t_k}^{t_{k+1}} \lambda(t) dt}
\end{eqnarray}
where the $\{t_k\}$ are the spike times and $\lambda(t)$ is the
history-dependent spiking rate from the CSM.  If the CSM describes the data
well, then rescaled ISI's $\{\tau_k\}$ should follow a uniform distribution.
This can be tested using either a Kolmogorov Smirnov test or by plotting the
empirical CDF of the rescaled times against the CDF of the uniform distribution
(Kolmogorov Smirnov or ``KS'' plot) \citep{Brown-time-rescaling}.

\begin{figure}[p]
\includegraphics{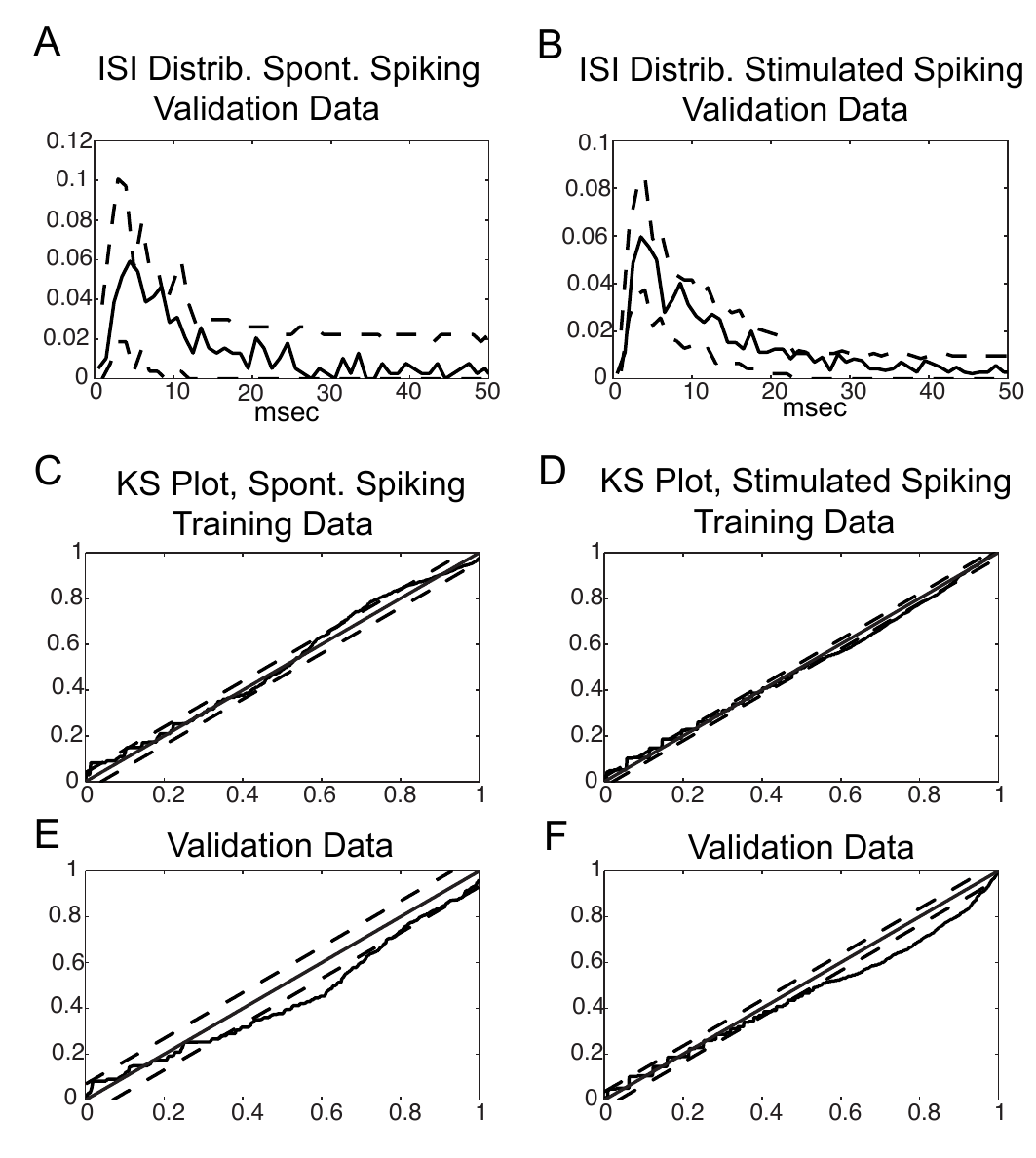}
\caption{Cross validation of CSMs reconstructed from spontaneously
firing and stimulus evoked rat barrel cortex on an independent 
validation training set.  (A,B) ISI distribution of spontaneously
and stimulus evoked firing validation sets and 99\% confidence bounds
bootstrapped from CSM.  (C-D)  Time rescaling plots of training
data sets for spontaneously firing and stimulus evoked firing respectively.
Dashed lines are
95\% confidence bounds and the solid line is the rescaled ISIs.  The
solid line along the digagonal is for visual comparison to an ideal fit.
(E-F) Similar time rescaling plots for the validation data sets.}
\label{fig:11}
\end{figure}

Figure \ref{fig:11} gives cross-validation results for the rat barrel cortex
neuron, during both spontaneous firing and periodic vibrissae deflection.  90
seconds of spontaneously firing spikes were split into a 75 second training set
and a 15 second validation set.  The 335 seconds of stimulus-evoked firing were
split into a 270 second training set and a 65 second validation set.  Panels
A and B show the ISI bootstrapping results for the spontaneous and stimulus
evoked firing respectively.  The dashed lines are 99\% confidence bounds from a
CSM reconstructed from the training set and the solid line is the ISI
distribution of the validation set.  The ISI distribution largely falls within
these bounds for both the spontaneous and stimulus evoked data.

Panels C-F display the time rescaling test.  Panels C and D show the time
rescaling plots for the spontaneous and stimulus evoked training data
respectively.  The dashed lines are 95\% confidence bounds.  The spontaneous KS
plot largely falls within the bounds.  The stimulus-evoked does not, but this
is expected because, as discussed, the CSM does not completely capture the
imposition of the external stimulus.  (The jagged ``steps'' in both plots
result from the 1 msec temporal discretization.)  Panels E and F show the time
rescaling plots for, respectively, the spontaneous and stimulus evoked
validation data.  The fits here are somewhat worse.  In the stimulated case,
this is not surprising.  In the spontaneous case the cause is likely
non-stationarity in the data, a problem shared with other spike train analysis
techniques, such as the Generalized Linear Model approaches described in the
next Appendix.  It should be emphasized that the point of reconstructing CSMs
is not to obtain perfect fits to the data, but instead to estimate the
structure inherent in the spike train, and the cross-validation results should
be viewed in this light.

\section{Causal State Transducers and Predictive State Representations}
\label{sec:CSTs-and-PSRs}

Mathematically, CSMs can be expanded to include the influence of external
stimuli on the process, yielding causal state {\em transducers}, which are
optimal representations of the history-dependent mapping from inputs to outputs
\cite[ch.\ 7]{CRS-thesis}.  Such causal state transducers are a type of
partially-observable Markov decision process, closely related to predictive
state representations (PSRs) \citep{predictive-representations-of-state}.  In
both formalisms, the right notion of ``state'' is a statistic, a measurable
function of the observable past of the process.  Causal states represent this
through an equivalence relation on the space of observable histories.  For
PSRs, the representation is through ``tests'', i.e., a distinguished set
input/output sequence pairs; the idea is that states can be uniquely
characterized by their probabilities of producing the output sequences
conditional on the input sequences.

An algorithm for reconstructing causal state transducers would begin by
estimating probability distributions of future histories conditioned on both
the history of the spikes {\em and} the history of an external covariate $Y$,
e.g.  $\Prob(X_{t+1}^{\infty}|X_{-\infty}^t,Y_{-\infty}^t)$, and otherwise be
entirely parallel to CSSR.  This has not yet implemented.

\end{appendix}

\end{document}